\documentclass[a4paper,11pt]{article}
\usepackage{timecenter}
\usepackage{times}

\usepackage{url}
\usepackage{alltt}
\usepackage{subfigure}
\usepackage{xspace}

\newcommand{\ie}{\textit{i.e.},\xspace}
\newcommand{\eg}{\textit{e.g.},\xspace}

\newcommand{\etc}{\textit{etc}.\@\xspace}

\begin{document}

\title{Efficient Management of Short-Lived Data}

\author{Albrecht Schmidt and Christian S.~Jensen}

\publication_history{May 2005, a \textsc{TimeCenter} Technical Report}

\trnumber{82}

\maketitle

\begin{abstract}

Motivated by the increasing prominence of loosely-coupled systems,
such as mobile and sensor networks, which are characterised by
intermittent connectivity and volatile data, we study the tagging of
data with so-called expiration times.  More specifically, when data
are inserted into a database, they may be tagged with time values
indicating when they expire, \ie~when they are regarded as stale or
invalid and thus are no longer considered part of the database.  In a
number of applications, expiration times are known and can be assigned
at insertion time.
We present data structures and algorithms for online management of
data tagged with expiration times.  The algorithms are based on fully
functional, persistent treaps, which are a combination of binary
search trees with respect to a primary attribute and heaps with
respect to a secondary attribute.  The primary attribute implements
primary keys, and the secondary attribute stores expiration times in a
minimum heap, thus keeping a priority queue of tuples to expire.  A
detailed and comprehensive experimental study demonstrates the
well-behavedness and scalability of the approach as well as its
efficiency with respect to a number of competitors.

\end{abstract}

\section{Introduction}
\label{sec:introduction}

We explore aspects of the implementation of an extension to Codd's
relational data model~\cite{Cod70}; the extension is a time-stamp
called \emph{expiration time} which is associated with each tuple in
the database.  By looking at a tuple's time-stamp, it is possible to
see when the tuple ceases to be part of the current state
of the database.
Specifically, assume that, when a tuple $r$ is inserted into the
database, it is tagged with an \emph{expiration time},
$t^\mathit{exp}(r)$.  Tuple $r$ is thus considered part of the current
state of the database from the time of insertion until
$t^\mathit{exp}(r)$.
Expiration-time semantics now ensures that operations, most
prominently queries, do not see tuples that have expired by the time
associated with a query.
Our study is motivated by the emergence and increasing prominence of
data management applications which involve data for which the
expiration time is known at
the time of insertion, updates are frequent, and the connectivity of
the data sources that issue the updates is intermittent~\cite{CFZ01}.
Applications which involve mobile networks, sensor networks, and the
Internet generally qualify as examples.

Data produced by sensors that measure continuous processes are often
short-lived.  Consider a sensor network of temperature sensors that
monitor a road network.  It may be assumed that a temperature
measurement is valid for at most a fixed number of minutes after it is
measured, or the duration of validity may be determined by more
advanced computations in the sensor network.  A central database
receives temperature measurements tagged with expiration times. A
measurement from a sensor then automatically disappears if the sensor
does not issue a new temperature measurement before the old
measurement expires.

While a temperature sensor network may be relatively static in nature,
mobile devices that frequently log on to and log off from access points
form a more dynamic network. In such a context, it is natural to tag
records that capture log-ons with expiration times so that a session
can be closed on the server side after a period of inactivity.  A
closely related scenario is the following: Consider stateless
protocols such as HTTP~\cite{W3C03} where cookies or session keys are
used to implement functionality like transactions and data
confidentiality.  A cookie or session key is naturally tagged with an
expiration time that indicates until when the key is considered valid.
Depending on the actual application, validity can range from values
below one second in, \eg~challenge-response protocols, to more than a
dozen minutes when used to implement session protocols on top of HTTP.

Another example application of expiration times is the monitoring of
availability of mobile devices through heartbeat tactics~\cite{BCK03}.
In this setting, mobile devices periodically emit heartbeat messages
to a central server.  If no heartbeat has been received from a device
for a specific duration of time, the device is assumed to be
unavailable (\eg~the device may have logged off).  A heartbeat message
may carry additional information such as the current position of the
device or a sensor value.  The advantage of heartbeats is that they
only incur half the network traffic of traditional ping/echo
availability.  Further attractive application areas for expiration
times include security models: in role-based security models, a role
often has a natural duration, after which the role and its associated
credentials expire.  Short-lived security primitives include one-time
passwords, session keys, challenges, Kerberos-tickets, and credentials
in challenge-response protocols~\cite{Sch96}.  Similarly, data in the
Domain Name System (DNS)~\cite{AL01} come with expiration times,
called time-to-live (TTL); they help to ensure that the translations
between domain names and IP addresses, both of which may change, are
up-to-date. 
By adding the notion of expiration time to a database language like
SQL, database management system (DBMS) designers can help application
programmers and software architects to simplify software architectures
and reduce code complexity while retaining the traditional transparent
semantics.  A user of an expiration time-enabled SQL engine needs not
be aware of the new concept, as expiration time $\infty$ can be assumed
for tuples for which no expiration time is provided explicitly.

A further benefit of the integration of expiration times into a DBMS
is that the number, and thus cost, of transactions especially in
distributed systems can often be reduced significantly because no
explicit delete statements need be issued; since transaction costs in
these settings are often an important bottleneck, overall system
performance can increase significantly.
We also find that the concept of expiration times can simplify the
application logic needed in dynamic environments.  Because the client
side needs not issue `clean-up' transactions to the server, code
complexity and the number of possible points of failure decrease.

The \emph{contributions of this paper} are as follows.  \emph{First},
motivated by the ubiquity of loosely-coupled distributed systems with
unstable connections such as mobile and sensor networks, we argue that
DBMS support for expiration time benefits applications, as pointed out
above.  Support for expiration time goes well with networks with
intermittent connectivity and volatile topologies, in that it reduces
the dependence on continuous, high-quality network connectivity.
expiration time.  \emph{Second}, we point out that the notion of data
expiration can be taken to the declarative level and be supported by
database query languages like SQL.  \emph{Third}, as the main
technical contribution, the paper presents online main-memory
algorithms and data structures that are capable of handling data
expiration efficiently; this implies that expired data are
automatically removed from the database without the need for user
interaction.  Through eager expiration policies, we achieve that data
items are removed from the database as early as possible; we thus
minimise the size of the database and enable timely trigger
support. \emph{Fourth}, a comprehensive experimental study offers
insight into resource consumption and other performance
characteristics such as scaling behaviour, response times, and
throughput.

The remainder of the paper is structured as follows. The next section
covers related work, and Section~\ref{sec:preliminaries} briefly
outlines the assumed extension to the relational model, thus offering
context for the subsequent sections.  Section~\ref{sec:treaps} defines
the notion of treaps and how to implement the most common operations.
Section~\ref{sec:experiments} presents the results of a comprehensive
evaluation of the performance characteristics of persistent treaps and
a comparative study of the performance of treaps with respect to
various competing data structures; it also covers a variety of
functional issues in relation to expiration times and the use of
treaps.  After a review of related work, the final section summarises
the paper and identifies promising directions for future research.

\section{Data Model Aspects}
\label{sec:preliminaries}

We assume the following basic setting for our research.  A number of
data sources exist which tag the values they emit with expiration
times. A relational view of these sources is provided, where only
valid tuples are exposed to queries. Valid tuples are those that have
not expired by the time a query is issued, \ie~the time associated
with the transaction in which the query is embedded~\cite{JL01}. Thus,
expiration times can be seen as a database-internal function
$t^\mathit{exp}:\textit{tuples}\to\textit{timestamps}$ from tuples to
timestamps.
Assuming that a database $\mathit{db}^\mathit{exp}$ of tuples with
expiration times is given and that the time a query $q$ is issued is
given by $\tau_q$ then the tuples seen by query $\mathit{q}$
is given as follows:
\[ \{\,r\, | \,r \in \mathit{db}^\mathit{exp} \wedge t^\mathit{exp}(r) >
\tau_q \} \]

Although our focus is on implementation issues and performance, it is
relevant to briefly cover data model aspects. Specifically, we provide
a basis for supporting expiration time by allowing the user to define
a table as \texttt{EXPIRABLE}. This affects statements to create and
alter tables as well as insertion and update commands.  An example
follows.

\begin{small}
\begin{alltt}
  CREATE \textit{EXPIRABLE} TABLE Session (
     SessionID CHAR (10),
     UserID VARCHAR (8), \dots );
\end{alltt}

\begin{alltt}
  INSERT INTO Session
  VALUES ('ABCDEFGHIJ', 'JOE24', \dots)
  \textit{EXPIRES} TIMESTAMP '2004-03-31 16:01:23';
\end{alltt}  
\end{small}

We assume that expiration times are implemented by a standard data
type like \texttt{TIMESTAMP}~\cite{MS93,SN99} and are expressed in
terms of the clock of the DBMS.
Note that an expirable table such as Session does not capture the
expiration times of its tuples using a regular, explicit column of
type \texttt{TIMESTAMP}, but rather records expiration times
implicitly. This design decision, of having expirable tables rather
than tables with special, expirable columns, emphasises that
expiration times cannot be referenced in queries.  Another
paper~\cite{SJ03} argues that expiration times can be handled
automatically by a query engine.

It is a fundamental decision to associate expiration times with
tuples.  Arguably, they could be associated with other constituents of
the relational model, including individual attribute values or even
attributes and other schema elements.  This design decision is
motivated by a desire for clear semantics, simplicity, and
practicality.

In the kinds of distributed environments that motivate this research,
tuples for insertion into the database along with their expiration
times are born in the network external to the database. When an
expiration time is first specified at a data source, this time is
likely to be specified in terms of a clock that is different from the
DBMS clock; and the expiration time may even be specified as a
duration relative to the time when the insertion containing the tuple
is issued.
When an expiration time is given using a clock that is different from
that of the DBMS, we assume that the translation between that clock
and the DBMS clock occurs before the tuple and expiration time are
inserted into the database, and we do not consider such translations
in this paper. We believe that such translations should occur in
middleware.

\section{Overview of Treaps}
\label{sec:treaps}

This section introduces persistent treaps for data expiration. Focus
is in the algorithms that relate to persistence and expiration.

\subsection{Introduction}
\label{sec:intro}

From a data structure point of view, a treap is a combination of a
\underline{tre}e and a he\underline{ap}.  With respect to a (primary)
key attribute, it is a binary search tree; with respect to a second,
non-key attribute, it is a heap.  The idea we elaborate on in the
remainder of the paper is to use the key attribute for indexing while
managing expiration times using the second non-key attribute.

Before we go into the details of the data structure, we define what we
mean by \emph{persistent}. We call a data structure (\textit{fully})
\textit{persistent} or (\textit{fully}) \textit{functional} if an
update to the data structure produces a new version without altering
the original.  This implies that we can both query and update old
versions; in this paper, we make use of the former to implement
concurrency but do not take advantage of the latter since this
functionality is orthogonal to what we desire.  In this sense, the
treaps in this paper can also be seen as \emph{partially persistent};
a data structure is called \emph{partially persistent} if old versions
can only be retrieved, but not updated~\cite{Oka98}.  We remark that
persistent data structures therefore lend themselves to enabling
concurrent access through versioning~\cite{BHG87}.  The main technical
advantage why we use persistency is that as long as only one thread
updates the treap, read-only access can be implemented with a minimum
of locking, which is desirable in a main-memory environment.

The structure of a treap node is shown in Figure~\ref{fig:datatypes}.
The layout of the tree is binary: each node has a left and a right
child, a key, and an expiration time; it also has a value field, which
may contain arbitrary data such as non-key attributes.
The basic setting just outlined of course impacts query processing, as
we point out next.
\begin{figure}[htp]
  \begin{tabbing}
    \quad\=\quad\=\quad\=\quad\=\quad\=\quad\=\quad\=\quad\=\quad\=\kill
    \>\textbf{class} Node;\\
    \>\textbf{class} Inner \textbf{extends} $<$ k, t, v $>$ Node \{\\
    \>\>\textit{left child}: Node;\\
    \>\>\textit{key}: k;\\
    \>\>\textit{expiration}: t;\\
    \>\>\textit{data}: v\\
    \>\>\textit{right child}: Node;\\
    \>\}~/* Instantiate with: Inner (lc, k, t, c, rc)  */\\
    \>\textbf{class} Leaf \textbf{extends} Node \{\};~/* Instantiate with: Leaf */
  \end{tabbing}
  \caption{Treap Node Datatype}
  \label{fig:datatypes}
\end{figure}

\subsection{Example}
\label{sec:example}

We proceed to exemplify how persistent treaps can be used to support
expiration times efficiently.  Focus lies on eager removal of expired
data.

\begin{figure}[htp]
  \centering
    \includegraphics[width=\textwidth]{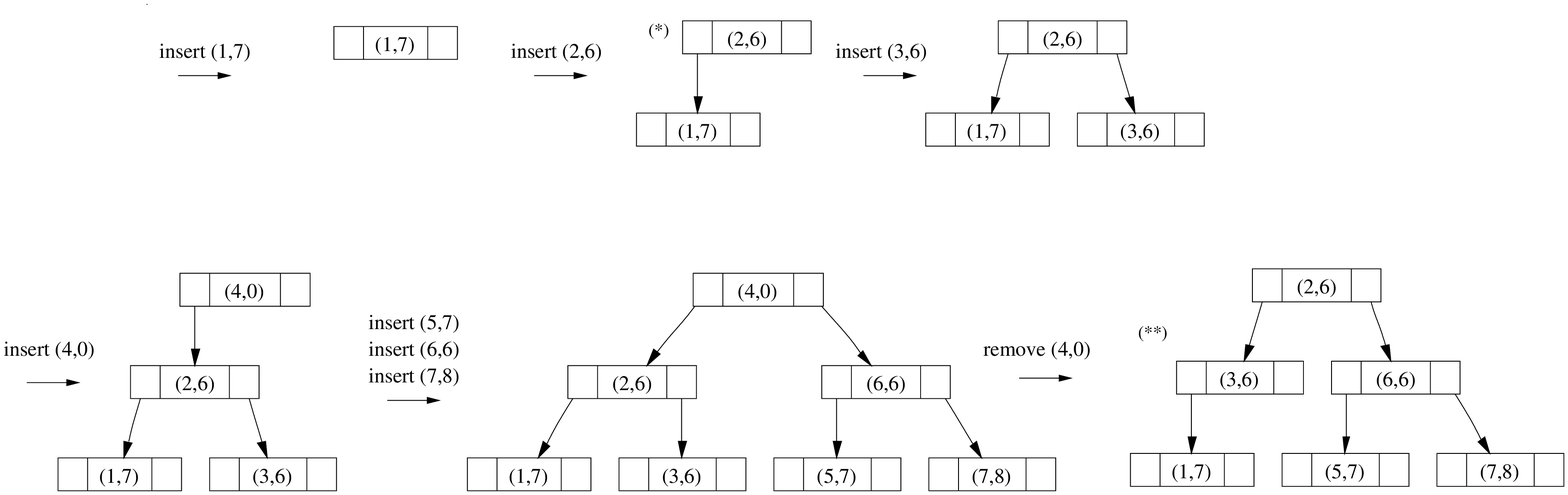}    
  \caption{Example Treap}
  \label{fig:example_full}
\end{figure}

Figure~\ref{fig:example_full} shows the construction of a treap given
the following sequence of key/expiration time-pairs to be inserted:
$$(1,7),(2,6),(3,6),(4,0),(5,7),(6,6),(7,8)$$
A pair $(k,e)$ denotes a tuple with key $k$ and associated expiration
time $e$.  Note that, for the time being, we assume that the key and
the expiration time are statistically independent; in
Section~\ref{sec:extensions}, we discuss what happens when we relax
this assumption; we now just remark that we can use a hash function on
the key to achieve independence.  The last step in
Figure~\ref{fig:example_full} consists of removing the root node from
the treap, \ie~carrying out an expiration, for example at time~1.  The
algorithms that do the actual work are discussed in the sequel.

For the time being, we have a quick look at the peculiarities of the
pseudo-code presented in this paper.  First, the persistence is
reflected in the code by the absence of the assignment operator and,
instead, the allocation of new objects with the \textbf{new} keyword
whenever an update is performed.  The function \textbf{new} $t~(a_1,
a_2, \dots, a_n)$ allocates a new object of type $t$ and initialises
it by calling the respective constructor with the arguments $a_1, a_2,
\dots, a_n.$

Second, extensive use of ML or Scala-style \emph{pattern
  matching}~\cite{Ode04} is made to bind parts of complex, nested data
structures to variables in a concise manner avoiding combinations of
nested if-statements.  For example, assuming the class definitions of
Figure~\ref{fig:datatypes}, using the second treap in the first row of
Figure~\ref{fig:example_full} (marked (*)) and the find function of
Figure~\ref{fig:findkey} then the first clause of
Figure~\ref{fig:findkey} is executed as follows.
Assume we want to find the node with the key $1$, \ie~we call `find
($\mathit{treap},1$)' where $\mathit{treap}$ is bound to (using the constructor
notation) `Inner (Inner (Leaf, 1, 7, $\bot$, Leaf), 2, 6, $\bot$, Leaf)'.  

\begin{figure}[htp]
  \begin{tabbing}
    \quad\=\quad\=\quad\=\quad\=\quad\=\quad\=\quad\=\quad\=\quad\=\kill
\scriptsize{1}\>\textbf{function} find (node, key) =\\
\scriptsize{2}\>\>\textbf{match} node \textbf{with}\\
\scriptsize{3}\>\>$|$ Inner (\_, k, \_, item, \_) \textbf{when} (key = k) $\to$ item\\
\scriptsize{4}\>\>$|$ Inner (left, k, \_, \_, right) $\to$\\
\scriptsize{5}\>\>\>\textbf{if} (key $<$ k)\\
\scriptsize{6}\>\>\>\textbf{then} find (left, key)\\
\scriptsize{7}\>\>\>\textbf{else} find (right, key)\\
\scriptsize{8}\>\>$|$ Leaf $\to$ \textbf{raise exception} (\textit{Key is not in treap})
  \end{tabbing}
  \caption{Lookup of a Primary Key}
  \label{fig:findkey}
\end{figure}

If we match against it the pattern `Inner (\_, k, \_, item, \_)
\textbf{when} (key = k)' (line 3, Figure~\ref{fig:findkey}), the
following variable bindings are created: $\mathrm{k}=2,
\textrm{item}=\bot$ ($\bot$ denotes a non-applicable variable in our
case, \ie~we do not use the data field in this example).  The
underscore `\_' in a constructor denotes a `don't care' variable that
is present in the class, but for which no binding is created.  Since
$\mathrm{k}$ is bound to $2$ and the function argument $\mathrm{key}$
to $1$, the \textbf{when} clause evaluates to false and the pattern
does not match.  However, the pattern in line 4 matches, and the
following bindings are created: $\mathrm{left}=\textrm{Inner (Leaf, 1,
  7, $\bot$, Leaf)}$, $\mathrm{right}=\textrm{Leaf}$, and
$\mathrm{k}=2$.  Since the $\mathbf{if}$ statement evaluates to true,
function find is called recursively and terminates successfully.

We recall that, conceptually, treaps are a combination of trees and
heaps.  With respect to the key $k$, they are binary search trees, and
in regard of the expiration time $e$, they are heaps.  If a treap is
well-balanced, \ie~structurally similar to a height-balanced binary
tree, the first property guarantees that we can execute look-up
queries in logarithmic time.  The second property implies that nodes
with minimal expiration times cluster at the treap root.  If the root
node has expired, \ie~its expiration timestamp $e$ is smaller than the
current time, we can simply remove it using the procedure described
next.  This is advantageous because we can keep the amount of stale
data to a minimum using an eager deletion policy: as long as the root
node is stale we remove it.  Since stale data cluster at the root, no
search is required.  Furthermore, this strategy has the advantage that
we essentially only need one procedure for both expiration and
deletion: indeed, expiration is implemented as a small wrapper around
the deletion algorithm (see~Figure~\ref{fig:expire}).

\subsection{Persistent Treaps in Detail}
\label{sec:treaps_detail}

This section introduces the most important operations on persistent
treaps.  Since we are not aware of any other work that presents
algorithms for the persistent variant of treaps, we describe the
functions that concern persistence in detail.

\subsubsection{Maintaining Balance}

\begin{figure}[htp]
  \centering
  \includegraphics[width=.8\textwidth]{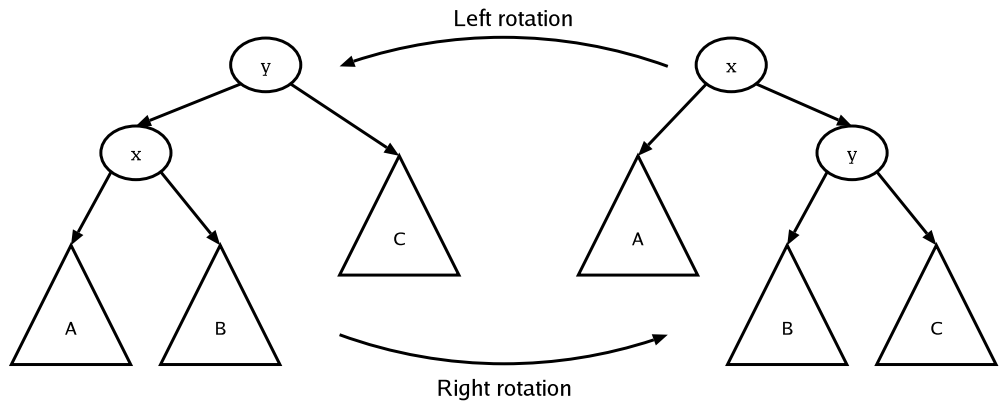}
  \caption{Left and Right Rotations}
  \label{fig:rotations}
\end{figure}

Like many other balanced tree structures, the insert and delete
functions of treaps maintain balance through order-preserving node
rotations.  Rotation, illustrated in Figure~\ref{fig:rotations}, is
done in two slightly different ways.

The insert function only rotates nodes on the path from a leaf, namely
the newly inserted node, to the root.  The delete function uses
rotations to move an interior node to the leaf level without violating
the order of the tree.  Due to the persistence property of our kind of
treap, rotations during inserts and deletes are implemented by
slightly different code: for insertion (Figure~\ref{fig:insertkey}),
the local function that implements the rotations is called
\textit{rebalance}; for deletion (see Figure~\ref{fig:expire}), it is
called \textit{percolate}.

\subsubsection{Insertion}

Insertion is a two-stage process.  First, we insert a pair $(k,e)$ as
if the persistent treap was a persistent binary tree on $k$.  We then
execute rotations to re-establish the heap property (while retaining
the binary tree property), which may have been violated.  Insertion
works as displayed in Figure~\ref{fig:insertkey}; it illustrated in
several places in Figure~\ref{fig:example_full}.

\begin{figure}[htbp]
  \begin{tabbing}
    ~~\=~~\=~~\=~~\=~~\=~~\=~~\=~~\=~~\=\kill
\textbf{function} insert (tree, key, time, item) =\\
\>\textbf{local function} rebalance (node) =\\
\>\>\textbf{match} node \textbf{with}\\
\>\>$|$ Inner (Inner (s1, u, t', i', s2), v, t, i, s3) \textbf{when} ($t > t')$ $\to$\\
\>\>\>\textbf{new} Inner (s1, u, t', i', \textbf{new} Inner (s2, v, t, i, s3))\\
\>\>$|$ Inner (s1, u, t, i, Inner (s2, v, t', i', s3)) \textbf{when} ($t > t')$ $\to$\\
\>\>\>\textbf{new} Inner (\textbf{new} Inner (s1, u, t, i, s2), v, t', i',s3)\\
\>\>$|$ \_ $\to$ node;\\
\>\textbf{match} tree \textbf{with}\\
\>$|$ Inner (\_, k, \_, \_, \_) \textbf{when} (key = k) $\to$ tree\\
\>$|$ Inner (left, k, t, i, right) $\to$\\
\>\>\textbf{if} (key $<$ k)\\
\>\>\textbf{then} rebalance (\\
\>\>~~~~~~\textbf{new} Inner (insert (left, key, time, item), k, t, i, right)\\
\>\>\textbf{else} rebalance (\\
\>\>~~~~~~\textbf{new} Inner (left, k, t, i, insert (right, key, time, item)))\\
\>$|$ Leaf $\to$ \textbf{new} Inner (Leaf, key, time, item, Leaf)
  \end{tabbing}
  \vspace*{-6pt}
  \caption{Insertion into Persistent Treaps}
  \label{fig:insertkey}
\end{figure}

The second phase allocates new memory as it re-establishes the heap
property.  This fact and because the function runs through the tree
twice (top to bottom for insertion and bottom to top for rebalancing)
may seem to make insertion a comparatively expensive operation;
however, since the first phase already populates the CPU caches with
the nodes needed in the second phase, the overhead is not too large.
The performance figure later in this paper quantify the cost of
insertion relative to expiration.  The amortised cost of insertion is
$O(\log n)$ time where $n$ is the number of elements stored in the
treap~\cite{SA96}.  Additionally, each insertion also allocates
$O(\log n)$ memory by producing a new version of the data structure;
however, since we use persistence to implement concurrency rather than
provide access to historical versions of the data, memory management
automatically reclaims $O(\log n)$ memory per insertion once it is not
used by other threads anymore.  Thus, for single-threaded applications
the overall memory requirements per insertion are not higher than for
non-persistent treaps.  In the case of multi-threaded applications,
old treap versions are reclaimed as soon the owning thread terminates.
For practical workloads, this usually implies that persistence does not
incur a memory overhead.

\begin{figure}[htp]
  \begin{tabbing}
    \quad\=\quad\=\quad\=\quad\=\quad\=\quad\=\quad\=\quad\=\quad\=\kill
\textbf{function} remove (node, key) =\\
\>\textbf{local function} percolate (node) =\\
\>\>\textbf{match} node \textbf{with}\\
\>\>$|$ Leaf $\to$ Leaf\\
\>\>$|$ Inner (Leaf, k, t, i, Leaf) $\to$ Leaf\\
\>\>$|$ Inner (Leaf, k, t, i, (Inner (lr, kr, tr, ir, rr))) $\to$\\
\>\>\>\textbf{new} Inner (lr, kr, tr, ir, percolate (\\
\>\>\>~~~~~~~~~~~~~~~~~~~~~~\textbf{new} Inner (Leaf, k, t, i, rr)))\\
\>\>$|$ Inner ((Inner (ll, kl, tl, il, rl), k, t, i, Leaf)) $\to$\\
\>\>\>\textbf{new} Inner (percolate (\\
\>\>\>~~~~~~~~~~~~~~~~~~~\textbf{new} Inner (ll, k, t, i, Leaf), kl, tl, il, rl))\\
\>\>$|$ Inner ((Inner (ll, kl, tl, il, rl) as left), k, t, i,\\
\>\>\>\>\>   ~(Inner (lr, kr, tr, ir, rr) as right)) $\to$\\
\>\>\>\textbf{if} (tl $\leq$ tr)\\
\>\>\>\textbf{then} \textbf{new} Inner (percolate (\\
\>\>\>~~~~~~~~~~~~\textbf{new} Inner (ll, k, t, i, rl)), kl, tl, il, right)\\
\>\>\>\textbf{else} \textbf{new} Inner (left, kr, tr, ir, percolate (\\
\>\>\>~~~~~~~~~~~~\textbf{new} Inner (lr, k, t, i, rr)));\\
\>\textbf{match} node \textbf{with}\\
\>$|$ Inner (\_, k, \_, \_, \_) \textbf{when} (key = k) $\to$ percolate (node)\\
\>$|$ Inner (left, k, t, i, right) $\to$\\
\>\>\>\textbf{if} (key $<$ k)\\
\>\>\>\textbf{then} \textbf{new} Inner (remove (left, key), k, t, i, right)\\
\>\>\>\textbf{else} \textbf{new} Inner (left, k, t, i, remove (right, key))\\
\>$|$ Leaf $\to$ \textbf{raise exception} (\textit{Key is not in treap}) \\[10pt]
\textbf{function} expire (node, time) = \\
\> \textbf{match} node \textbf{with}\\
\>$|$ Inner (\_, k, t, \_, \_) \textbf{when} (t $\leq$ time) $\to$\\
\>~~~~~~~~~~~~~~~~~~~~~~~~~~~~~~~~~~~~~~~expire (remove node k) time\\
\>$|$ \_  $\to$ node
  \end{tabbing}    
  \vspace*{-6pt}
  \caption{Removal and Expiration for Persistent Treaps}
  \label{fig:expire}
\end{figure}

\subsubsection{Removal and Expiration}

Like insertion, removal is a two-stage process.  The first step
consists of locating the node that contains a given key.  The second
step includes executing rotations so that the node sifts down and
eventually becomes a leaf.  After this has happened, it is simply
discarded.  Removal of a key is also covered in
Figure~\ref{fig:example_full} (marked (**) in
Figure~\ref{fig:example_full}).
The removal algorithm is shown in Figure~\ref{fig:expire} together
with the algorithm for expiration, which is simply implemented as a
deletion of the root element if its timestamp indicates that it is
stale.  We call the expiration function periodically to make sure that
the tree contains only a minimum of stale data; more advanced policies
are certainly possible and often appropriate. 
Since the remove algorithm returns a new version of the treap just
like insert, the discussion of resource requirements is similar to the
discussion of insertion.

\subsubsection{Other Operations}

Depending on the area of application, other operations on treaps make
sense as well.  For example, we can use full traversals of a treap to
create snapshots of the current state of the database for statistics,
billing, {\etc} Furthermore, if the less-than relationship between
keys returns sensible values, range queries on keys can be used to
quickly extract ordered intervals from the indexed keys.  These
operations are implemented exactly as for binary trees, so no code is
provided here.  However, a performance evaluation of full traversals
is presented in the next section.

\subsection{Concurrency}
\label{sec:concurrency}

Concurrency is achieved using
\textit{versioning}~\cite{BHG87}; it is implemented using the
node-copying method~\cite{DSD+89}.  Thus, only one thread is allowed
to update the data structure, but any number of threads can read from
it.  As pointed out earlier, this type of design pattern can be
implemented in a nearly locking-free manner and provides for
concurrent operations at the cost of increased memory allocation and
deallocation but not increased overall memory usage.  Modern
generational garbage collectors~\cite{CH97,DTM95} are optimised for
this kind of allocation pattern and provide favourable performance.
Despite the increased memory allocation and deallocation activity the
overall storage requirements are asymptotically not higher than
traditional single-version implementation (assuming a `standard'
database setting with a finite number of threads all of which feature
finite running times).

\section{Experiments and Evaluation}
\label{sec:experiments}

This section reports on empirical studies of the performance of
persistent treaps.

\subsection{Experimental Setup}
\label{sec:setup}

The experiments were carried out on a PC running Gentoo Linux on an
Intel Pentium IV processor at 1.5 GHz featuring 512 MB of main memory
available; no hard disk was used during the experiments. The CPU
caches comprise 8 KB at level 1 and 512 KB at level 2.  The compiler
used was gcc/g++ 3.2.  The performance data from which the graphs
displayed in this section were synthesised were gathered from
experiments lasting over 42 hours of runtime on a single machine.

\begin{figure}[htp]
  \centering
  \includegraphics[width=\columnwidth]{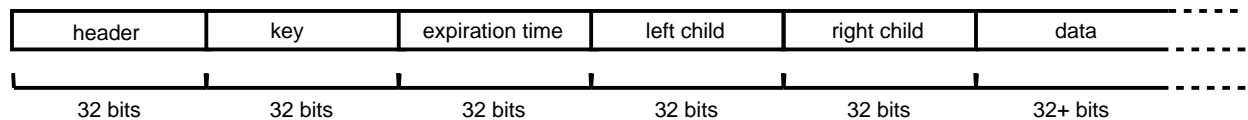}
  \caption{Physical Layout of an Internal Treap Node}
  \label{fig:node_layout}
\end{figure}

Figure~\ref{fig:node_layout} displays the physical layout of an
internal treap node in our implementation.
We fixed the size of the data field to 32 bits for our experiments.
All relevant data are inlined, so to access a key or expiration time,
we do not have follow a pointer, but we can read it locally in the
record.  This has been done mainly to improve cache
utilisation~\cite{ADH02}; in general, however, the data field may
contain a pointer to non-local data.
Furthermore, we are only concerned with the removal of data from the
treap.  Should the data section in Figure~\ref{fig:node_layout}
contain a pointer to a heap space or a disk block, taking care of this
is delegated to the application or run-time system.  How to do the
actual deletion in this case is an orthogonal matter: the literature
offers solutions in a secondary memory setting~\cite{SLM93}.

In order to explore the full potential and the limitations of
persistent treaps, we generated synthetic data to get the data volume
needed to test the behaviour of treaps in the limit.  The sensor and
network hardware available to us are unable to deliver the data
volumes necessary to determine the performance limitations of the data
structure.

\begin{figure}[htbp]
  \newcommand{\scale}{0.59}
  \centering
  \includegraphics[width=\scale\columnwidth]{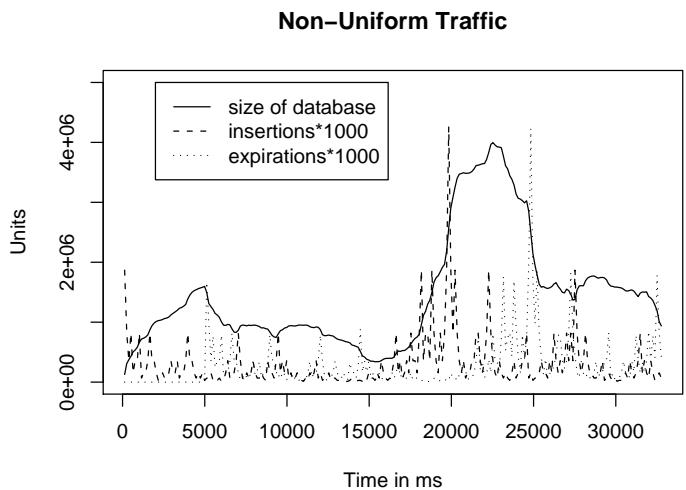}
  \vspace*{-12pt}
  \caption{Database Size and Operation for Non-Uniform Traffic}
  \label{fig:nonuni_perf}
\end{figure}

Figure~\ref{fig:nonuni_perf} exemplifies a workload we used.
The dashed line indicates the numbers of tuples that arrive at each
particular point in time.  For example, the peak at approx.~20,000
milliseconds denotes that 4,000 tuples arrive during the respective
interval and have to be inserted into the treap.  Without support for
expiration time, the network traffic would approximately double, and
each spike indicating the arrival of new data would be followed by a
spike indicating the deletion of the very data comprising the first
spike (assuming that all data expire a fixed duration after their
insertion).

We use the B-Model data generator proposed by Wang et
al.~\cite{WAF02,WCP02}, which is well suited for our purposes.  This
generator is capable of generating workloads while consuming only a
fraction of available system resources. Thus, the generator provides
enough performance not to flaw results.
To make it fit our purposes, we extend the generator to work with four
input parameters rather than the original three parameters. The three
original parameters, $b$, $l$, and $N$, are the bias, the aggregation
level, and the total volume, respectively; we refer to this model as
$\mathrm{BModel}(b,l,N)$.  The bias $b$ describes the roughness of the
traffic, \ie~how irregular it is and how pronounced the peaks are.
The aggregation level $l$ measures the resolution at which we observe
the traffic.  The parameter $N$ equals the sum of all measurements and
specifies the total amount of traffic.  The new, fourth parameter is a
random variable describing the distribution of the expiration times of
arriving network traffic, \ie~the time interval we consider the
$\mathrm{BModel}(b,l,N)$ arriving items valid.

\begin{figure*}[htp]
  \newcommand{\scale}{0.49}
  \centering
  \subfigure[Throughput under Uniform Traffic (b=0.5), 4 M Tuples]{\label{fig:large_ops}\includegraphics[width=\scale\columnwidth]{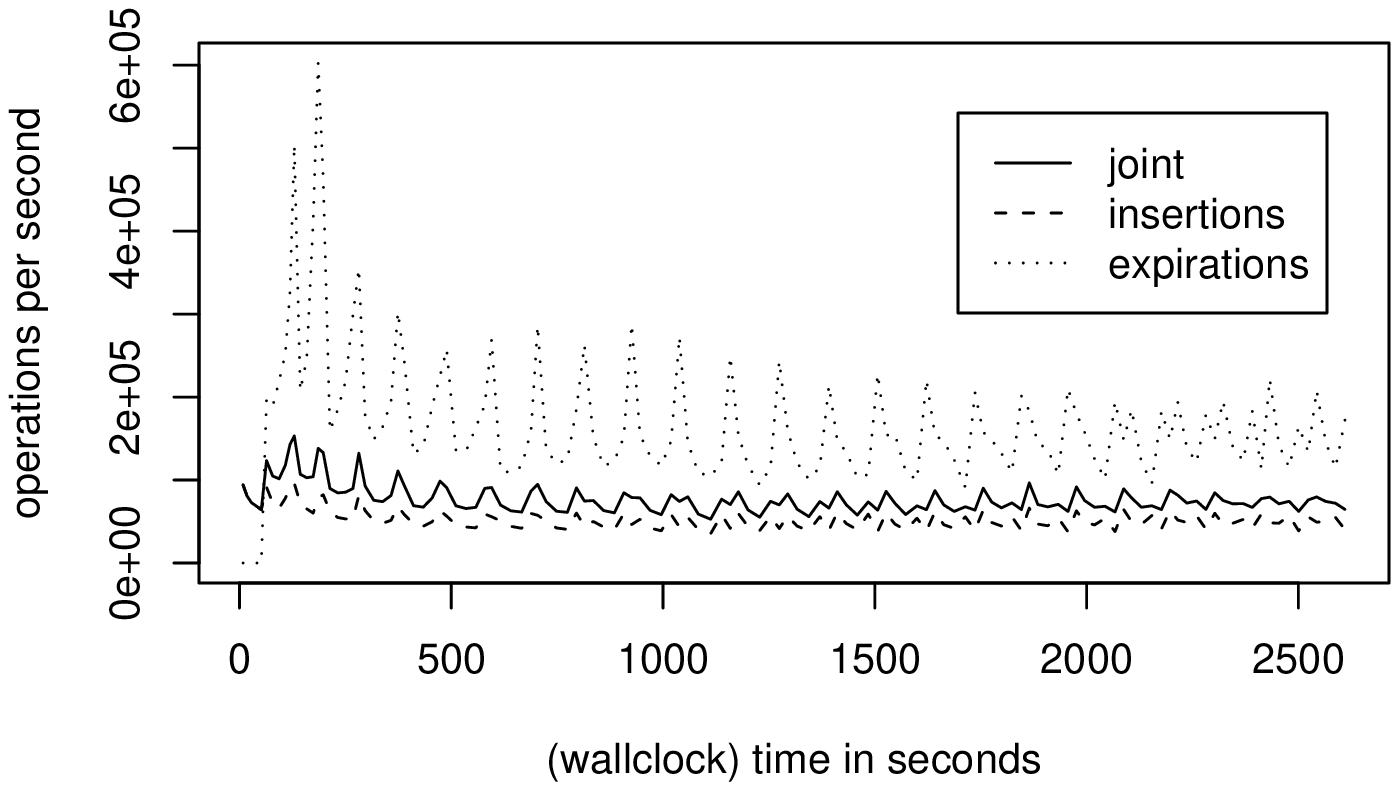}}
  \subfigure[Probability Density Functions for Uniform Traffic (b=0.5), 4 M Tuples]{\label{fig:large_ana}\includegraphics[width=\scale\columnwidth]{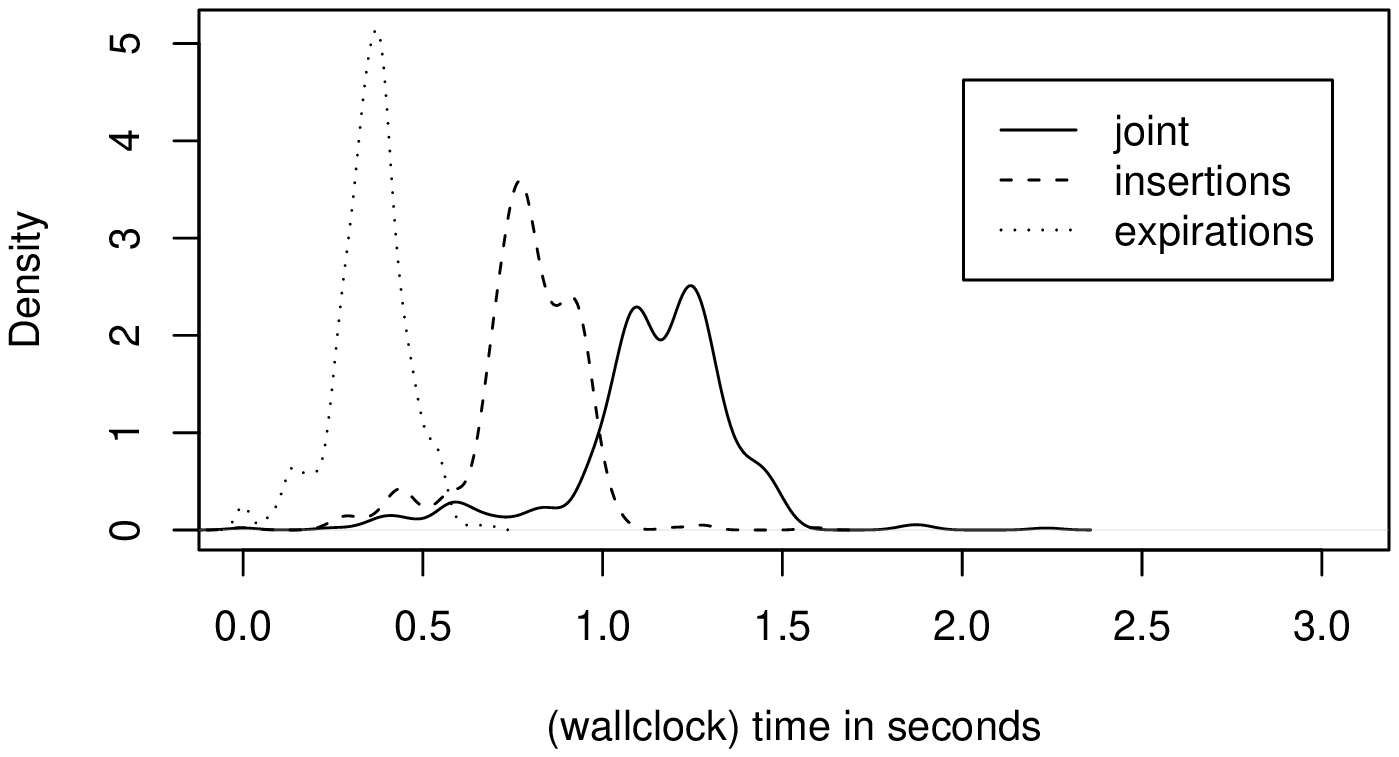}}    
  \subfigure[Throughput under Uniform Traffic (b=0.5), 40,000 tuples]{\label{fig:small_ops}\includegraphics[width=\scale\columnwidth]{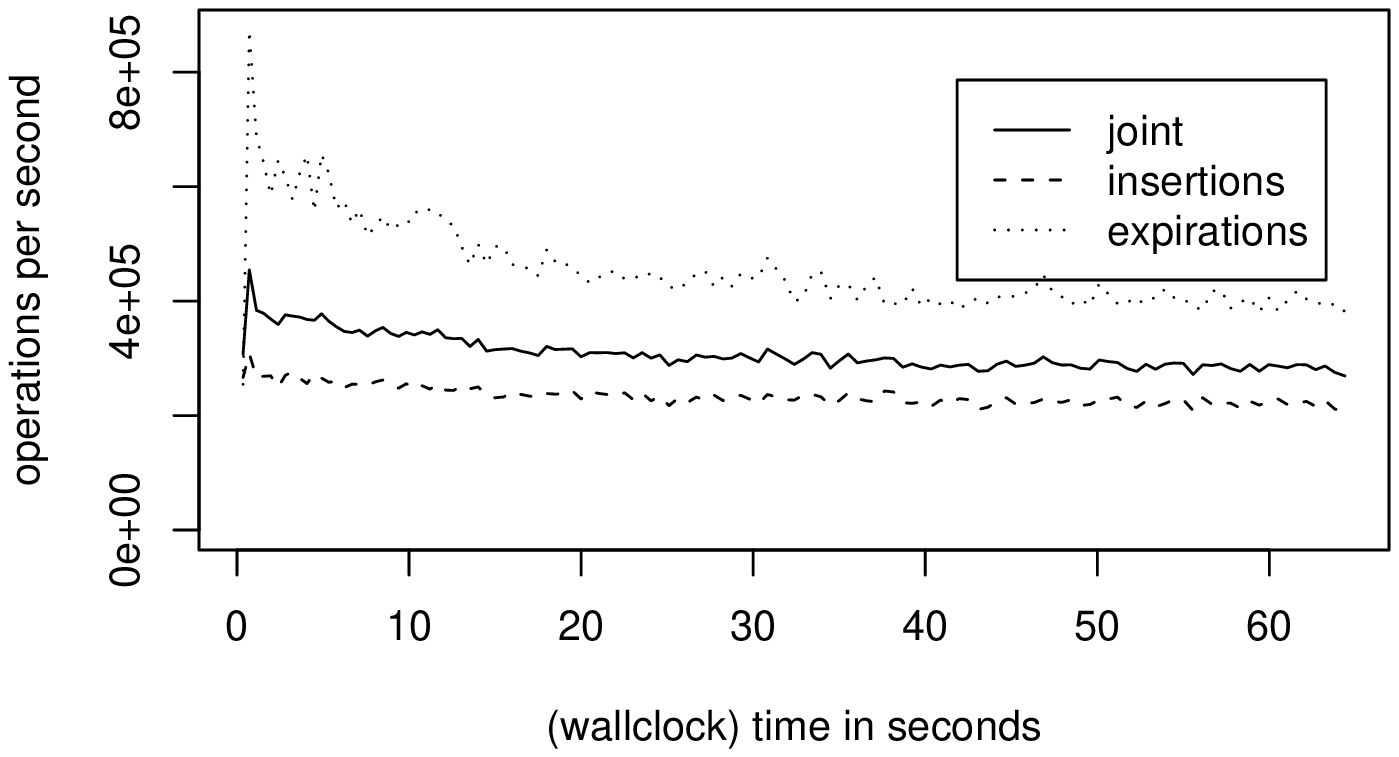}}
  \subfigure[Probability Density Functions with Uniform Traffic (b=0.5), 40,000 tuples]{\label{fig:small_ana}\includegraphics[width=\scale\columnwidth]{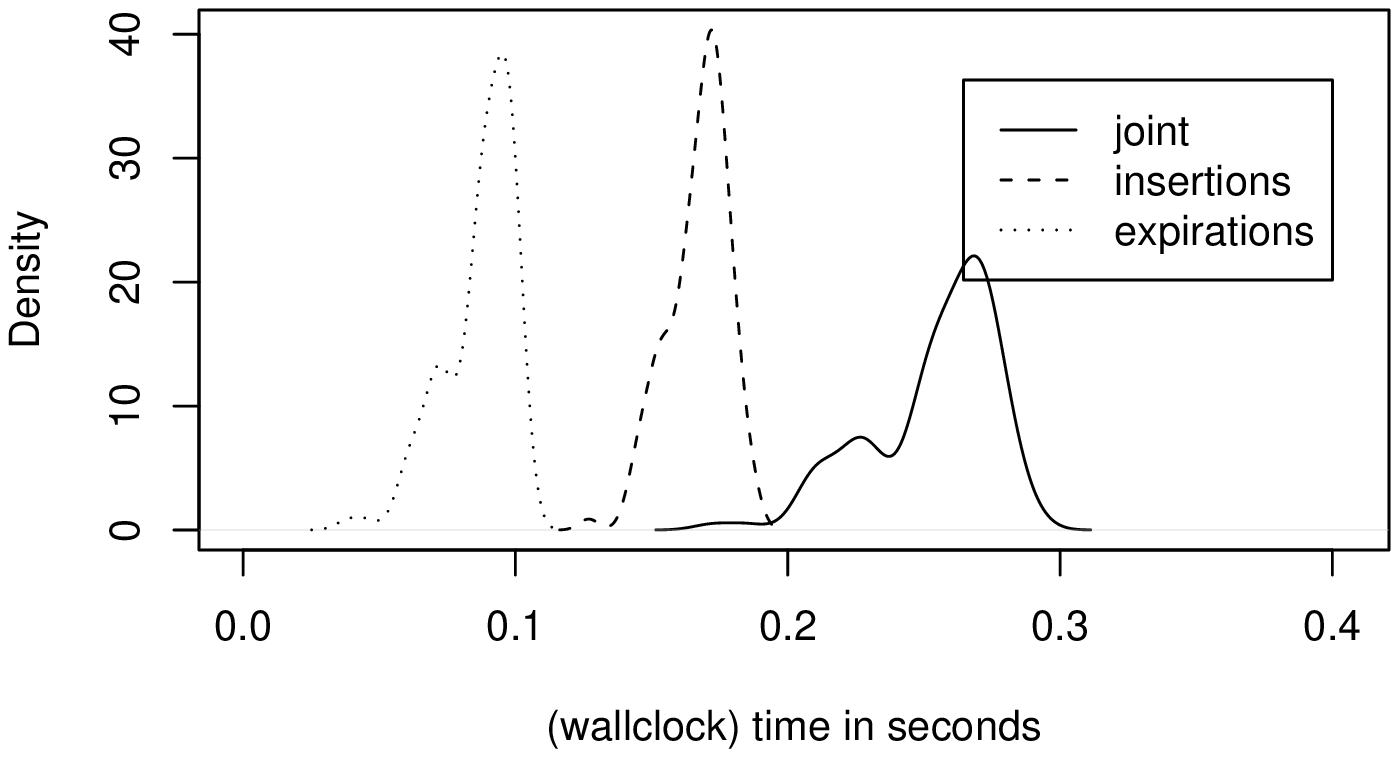}}
  \subfigure[Resource Utilisation for Run Displayed in Figure~\ref{fig:large_ops}]{\label{fig:utilisation_large}\includegraphics[width=\scale\columnwidth]{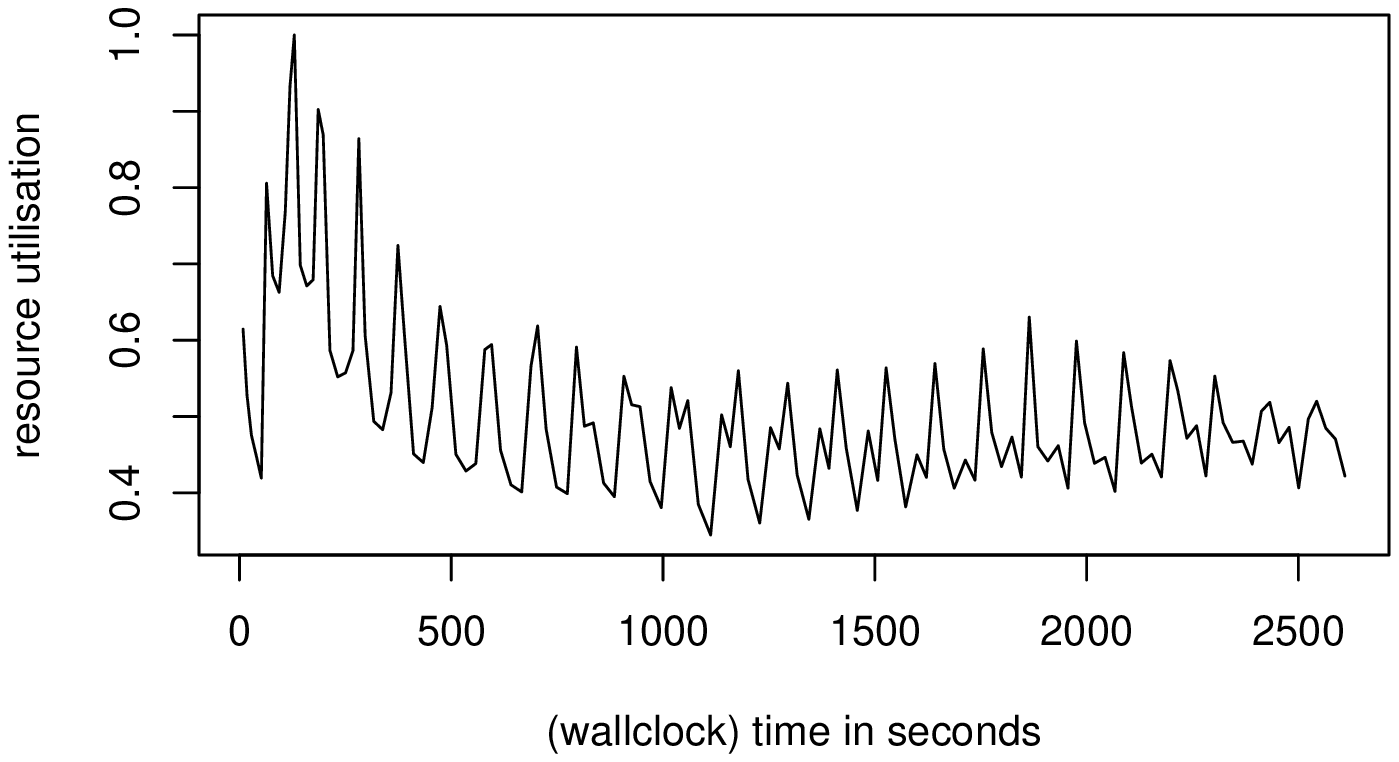}}
  \subfigure[Resource Utilisation for Run Displayed in Figure~\ref{fig:small_ops}]{\label{fig:utilisation_small}\includegraphics[width=\scale\columnwidth]{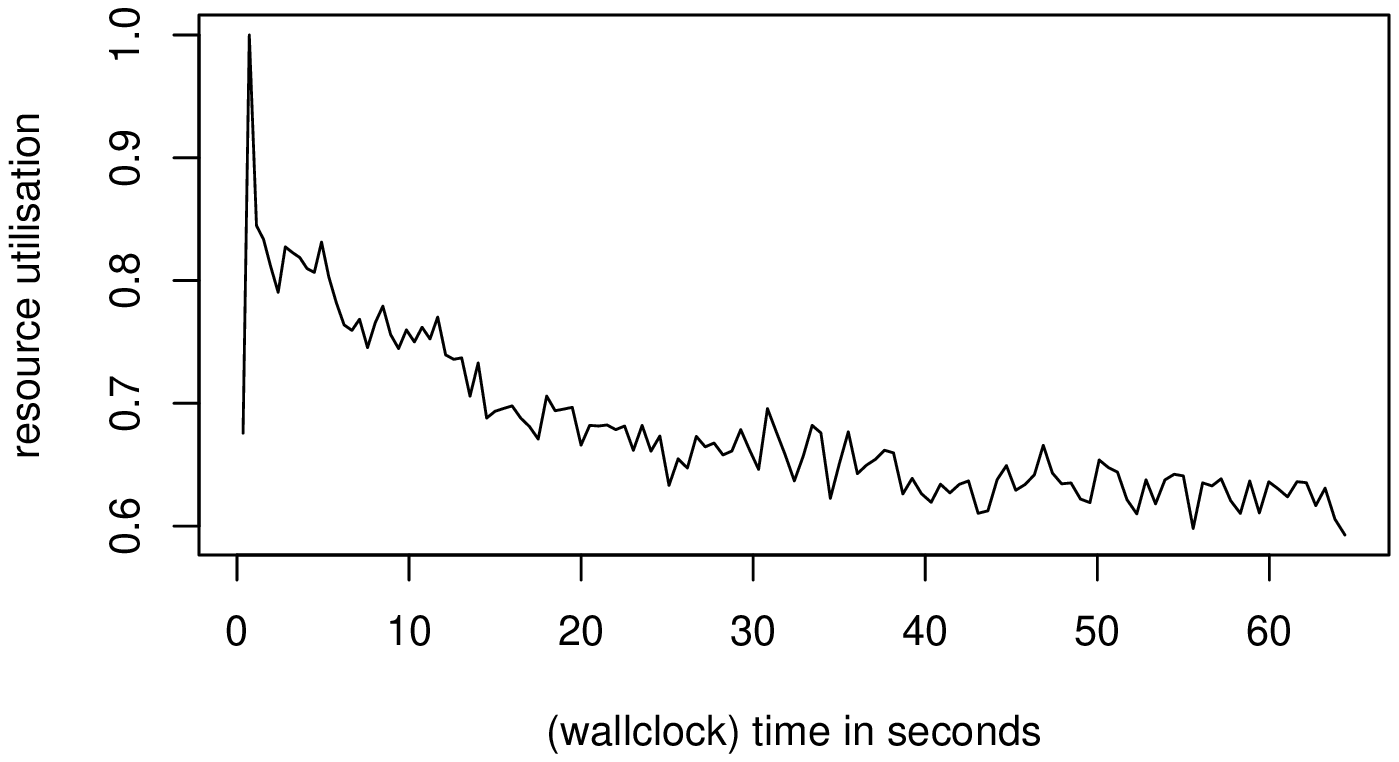}}
  \caption{Performance Impressions, Resource Consumption, and Peak Performance}
\end{figure*}

To get an impression of both maximum throughput and response to
extremely bursty traffic, we divided our experiments into two parts.
We first consider uniform traffic, \ie~$\mathrm{BModel}(0.5,l,N)$. This
is done to capture how treaps respond to continuous high workloads.
Since the versioning semantics call for frequent allocations of
memory, we can expect efficient memory management to be a key factor.

Next, we consider bursty traffic, \ie~$\mathrm{BModel}(b,l,N)$, $b \in
\; ]0.5,1.0[$. This is done to estimate how well treaps act under
workloads with more or less pronounced peaks. In these settings,
minimum and maximum throughput are of interest. Examining treaps in
this context is a first step towards the consideration of stochastic
quality-of-service guarantees.  For experiments which try to
illustrate scaling behaviour, $N$ is the parameter used to generate
databases of different size.  However, when we talk about the size of
a database, \eg~about 4 M tuples in Figure~\ref{fig:large_ops}, we
mean the average number of unexpired tuples residing in the database,
potentially after some bootstrap.

\subsection{Discussion of Treap Performance}
\label{sec:discussion}

We now turn our attention to
Figures~\ref{fig:large_ops}~through~\ref{fig:traversal}, which describe
the performance of the data structure under different stress patterns
and for different workloads.  We first investigate the performance of
updates; then we turn our attention to querying.

\subsubsection{Insertions and Expirations}

We first examine the behaviour of treaps under a uniform workload with
insertions into a large four million tuple database, \ie~after an
initial bulk load, the database consists of four million tuples on
average, with insertions and expirations basically cancelling out each
other.

Figure~\ref{fig:large_ops} shows the throughput for such a setting.
The conspicuous peaks are mainly due to comparatively cheap memory
allocation cost after major garbage collections.
Notice that the dotted line representing expiration remains, once
expirations set in, above the dashed line representing insertions;
thus, expirations are cheaper than insertions for large databases.
This reflects the structure of the insertion algorithm, requiring to
traversal from the root to a leaf and back.  On the other hand,
expiring the root only requires sifting the node to the leaf level
before discarding it.  Thus, expirations also require fewer memory
allocations than insertions.

Since treaps only guarantee amortised performance, it is also
interesting to learn to what degree the costs of the individual
operations differ. Due to high throughput -- sometimes more than
100,000 operations in our case -- it is very hard to monitor the cost
of an atomic operation without influencing the result to a degree that
renders it unusable.  Therefore, we move to a higher level of
aggregation and look at the length of intervals containing a fixed
number of operations, \ie~insertions and expirations.  This number was
fixed to 80,000 for the experiments.  While this rather large number
theoretically may obscure the variance in cost of individual
operations, we did not experience this problem and found it a good
compromise between unobtrusiveness and intuition.

In Figure~\ref{fig:large_ana}, the Probability Density Function
(PDF) of such an analysis displayed. 
We note that the local maximum of the solid line representing
expirations at $0.0$ indicates that the system wants to expire data,
but there is no stale data present. This results in an operation with
nearly zero cost.  Figure~\ref{fig:large_ana} also shows that
expirations are cheaper than insertions by a factor of approx.~two and
that the overall cost of a joint operation, insertions and expirations
combined, is reasonable.  Disregarding the phantom expirations of
nearly-zero cost, one can assume that the execution time of 80,000
insertions lies between about 1.0 and 1.4 seconds.  This suggests that
treaps behave reliably and predictably in practise.

To provide some evidence that the performance peaks in
Figure~\ref{fig:large_ops} are indeed memory management-related, we
concentrate on the results reported in Figure~\ref{fig:small_ops}.
This time, the database contains 40,000 tuples on average.  Now, the
local maxima in the graph are noticeably less pronounced than in the
large database.  It turns out that because of the small size, major
re-organisations of the storage space can be avoided, keeping the cost
of individual memory allocations on about the same level. This is
also reflected in the PDF displayed in Figure~\ref{fig:small_ana}.
The bandwidths of insertions, expirations, and of the joint PDF are
smaller in both absolute and relative terms.

Figures~\ref{fig:utilisation_large} and~\ref{fig:utilisation_small}
concern resource utilisation.
It can be seen that most of the time, the treap is able to insert data
at about half the maximum possible rate and that memory management
causes some pronounced spikes.  Again, for smaller database the spikes
remain less pronounced.

Uniform traffic can be considered the worst case for treaps in the
sense that it always has to deliver as much performance as possible.
In the case of non-uniform traffic, we can expect the system to
consume few resources when there are few operations, while running
resource-intensively when the numbers of operations peak.  This is also
demonstrated in Figure~\ref{fig:nonuni_perf}.  The straight line
indicates the database size.  Since we use an eager expiration and
removal policy, the line also reflects the number of valid,
\ie~non-expired, tuples in the database.
A note on the choice of the parameter $b=0.695234$ for non-uniform
traffic: we chose $b$ in this range because it is typical for Web
traffic~\cite{WCP02}, a scenario which is probably closest to our area
of application.

\subsubsection{Retrieval}

\begin{figure*}[t]
  \begin{center}
    \newcommand{\scale}{0.49}
    \subfigure[Scaling of 100,000 Uncorrelated Lookups for AVL Trees, Red-Black Trees, and Treaps.]{\label{fig:lookups}\includegraphics[width=\scale\columnwidth]{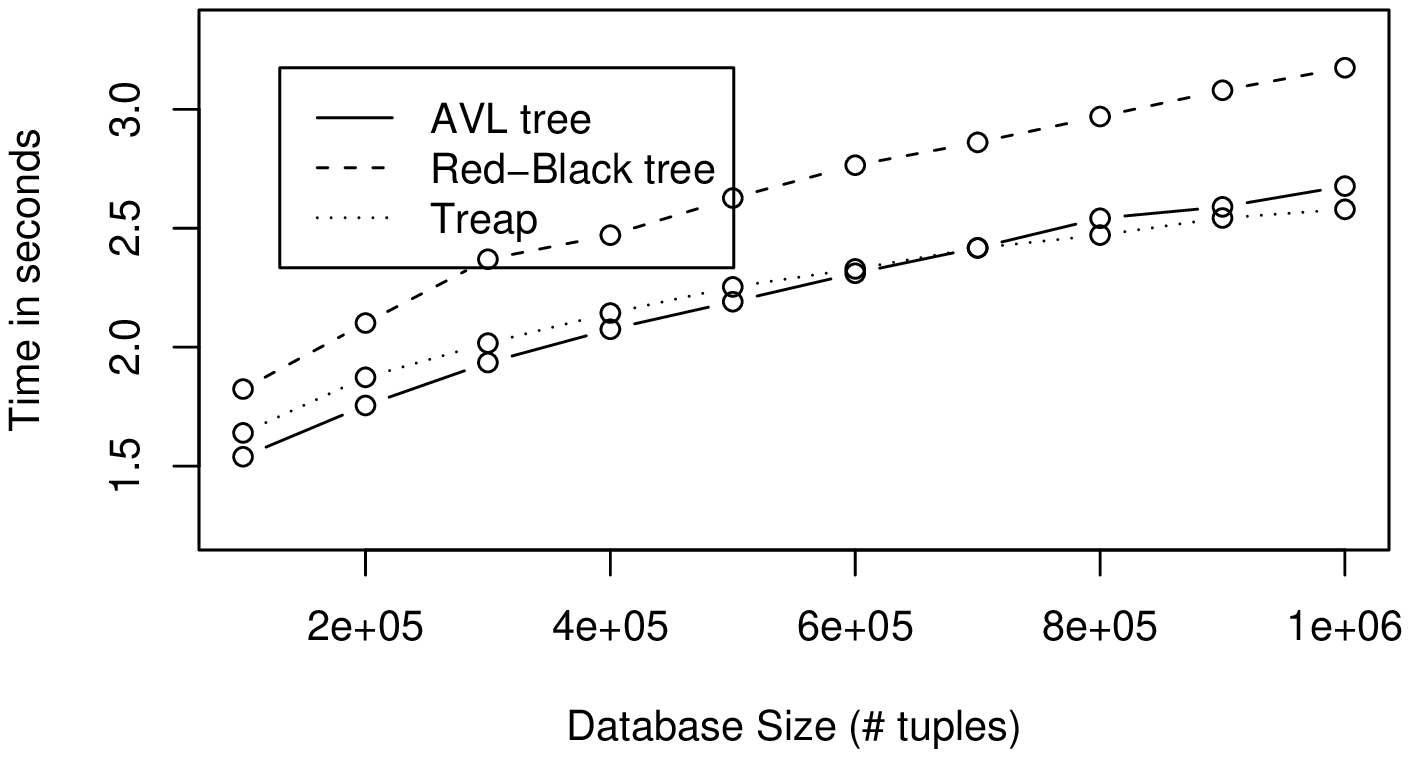}}
    \subfigure[Scaling of Complete In-Order Traversal of AVL Trees, Red-Black Trees, and Treaps.]{\label{fig:traversal}\includegraphics[width=\scale\columnwidth]{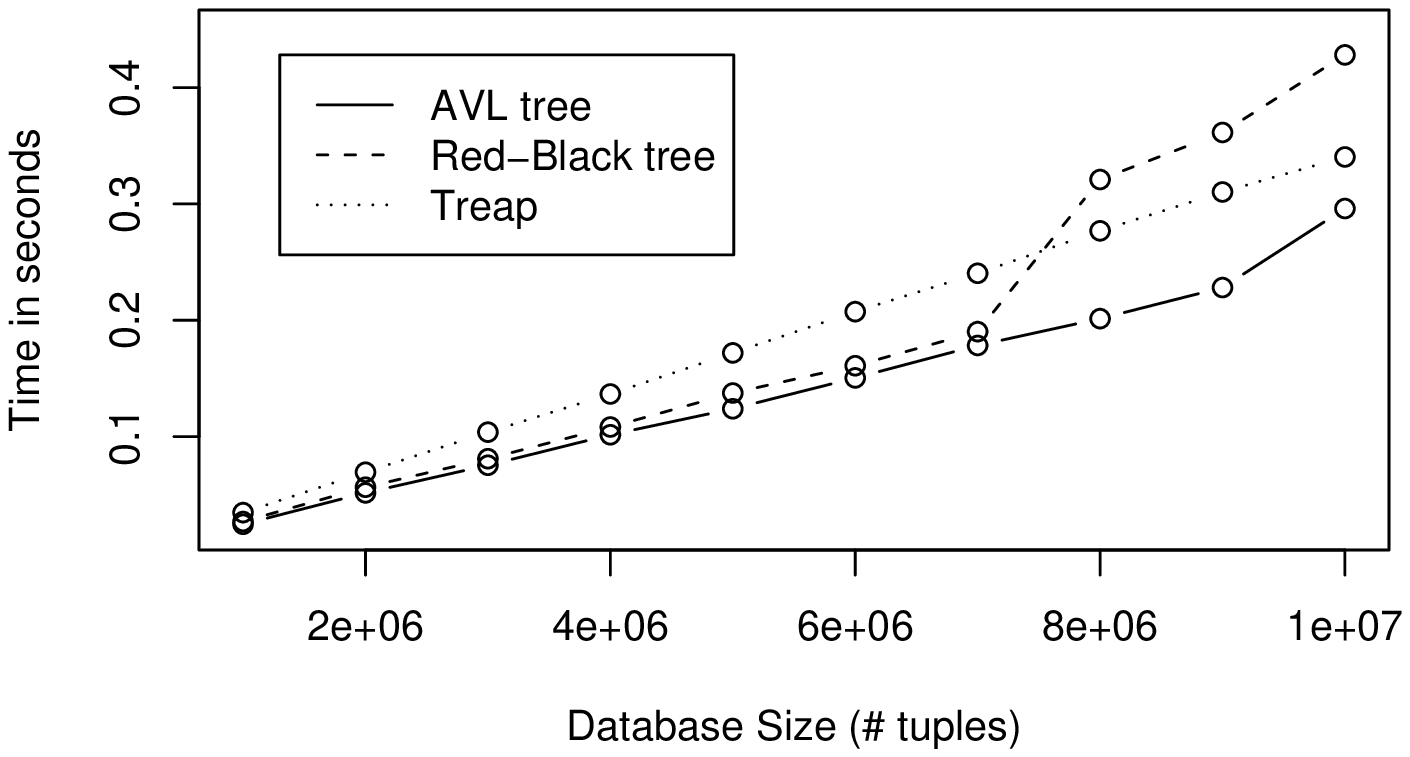}}
  \end{center}
  \caption{Lookup and Traversal in Database Containing Up To 16 M Tuples}
  \label{fig:queries}
\end{figure*}

Moving on to retrieval performance, Figure~\ref{fig:lookups} presents
the cost profile of uncorrelated lookups while varying the database
size.
The graph shows that the number of lookups per second decreases as the
database size grows. The graph is consistent with $O(\log n)$ key
lookup complexity.

Figure~\ref{fig:traversal} illustrates how expensive it is to traverse
a treap in an in-order fashion.
Traversal is an interesting operation, as it can be used for creating
snapshots, computing joins, {\etc} The operation is linear in the size
of the database, but benefits from caching: the path from the treap
root the current node is very likely to be resident in the cache
hierarchy.  Thus, the operation is surprisingly fast; traversing a one
million tuple database takes about one-third of a second on our test
platform.  This indicates that the versioning semantics of our treaps
does not impede full traversals since the number of arriving tuples is
certainly limited.  Figure~\ref{fig:queries} also displays the
performance of the same operations on AVL trees and Red-Black trees,
which are what we compare treaps against in the following subsection.

\subsection{Comparing Treaps to Competitors}
\label{sec:comparison}

To estimate the performance and resource consumption of the treap
index relative to other data structures, we compared the behaviour of
treaps to a number of competing approaches.  We use the following
methodology:
Besides requiring appropriate competitors to provide an index on the
key attribute, we distinguish between structures which support
\textit{eager} expiration and structures which do not.  Eager
expiration implies that we can remove expired data in a timely fashion
from the data structure so that, for example, ON EXPIRATION triggers
can fire as soon as the item becomes stale and not at some arbitrary,
later point in time.  Thus, eager expiration calls for priority
queue-like access to the data in addition to the index on the key
values.  We achieved this by combining the index structures with
priority queues.  Since our context requires us to work with
main-memory data structures, we chose AVL trees and Red-Black
trees~\cite{Knu98} as competitors to treaps.  To support expiration on
these structures, we applied (1)~periodic cleansing strategies, and
(2)~priority queue-supported, eager expiration strategies to both data
structures.  Since treaps may require us to apply a hash function to
key values and, thus, may not support range queries under certain
circumstances, we also compared the performance of treaps to
main-memory hash tables~\cite{Knu98}.  Again, we use plain hash tables
as well as heap-supported hash tables for eager expiration.

\subsubsection{Maintenance Costs}
\label{sec:maintenance}

To measure how dynamic a database instance is at a given point in time
we look at how many tuples of a snapshot would expire during a given
interval.  Formally, we introduce the notion of \textit{Rate of
  Expiration} (RoE), which is defined as
$$\textit{RoE}(d,u)=\mathit{\frac{\textit{\# expirations}}{\textit{\#
      live data}+\textit{\# expirations}}},$$
where $d$ is a database
containing at least one tuple and $u$ is an appropriate time interval;
for our experiments, $u$ is always the average lifetime of the
database tuples.
Thus, we look at a snapshot of a database the beginning of $u$ and at
the end of $u$ and compute the ratio of expiration during $u$ and the
sum of expirations and current data at the end of $u$.  Thus, the Rate of
Expiration is number between 0 and 1 (or 0\% and 100\%) which captures
how dynamic or how static a particular database state is by relating
the number of tuples expirations in a given time interval to the size
of the database.
Note that the RoE does not take
into account insertions and expiration from insertions; it only
measures the decay of a database state.  An RoE of 100\% would imply
that, during the interval $d$, all data expire, whereas an RoE of 0\%
implies that there are no expirations.
We note that expiration time-enabled data structures in general appear
particularly useful when $\textit{RoE}$ is relatively low, \ie~a
significant part of the database does not expire in the interval of
interest; high RoEs imply that the we have to dispose of large parts
of a database, which in turn implies that we have to scan the majority
of the data, which we can do anyway without supporting data
structures.

\begin{figure*}[!t]
  \newcommand{\scale}{0.49}
  \centering
  \subfigure[RoE=100\%]{\includegraphics[width=\scale\columnwidth]{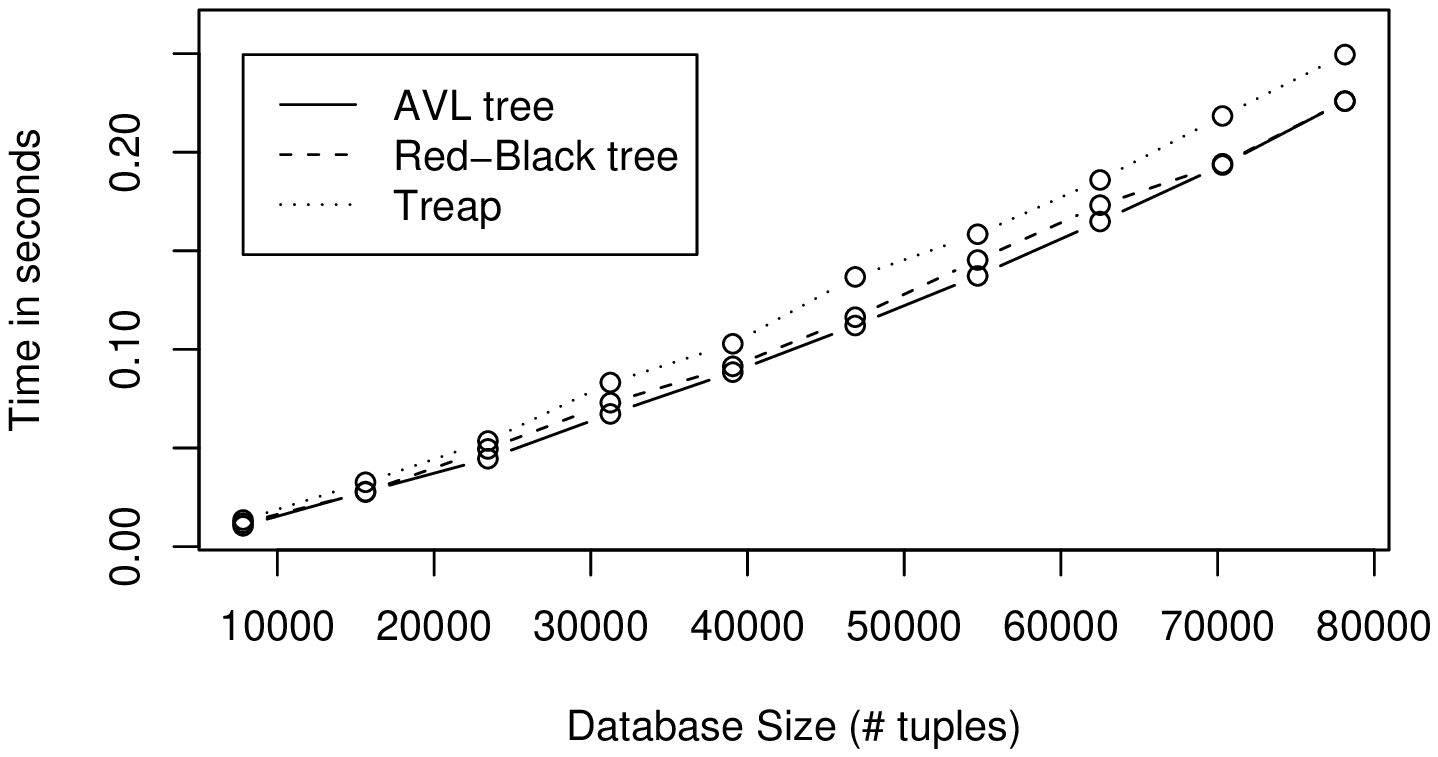}}
  \subfigure[RoE=5\%]{\includegraphics[width=\scale\columnwidth]{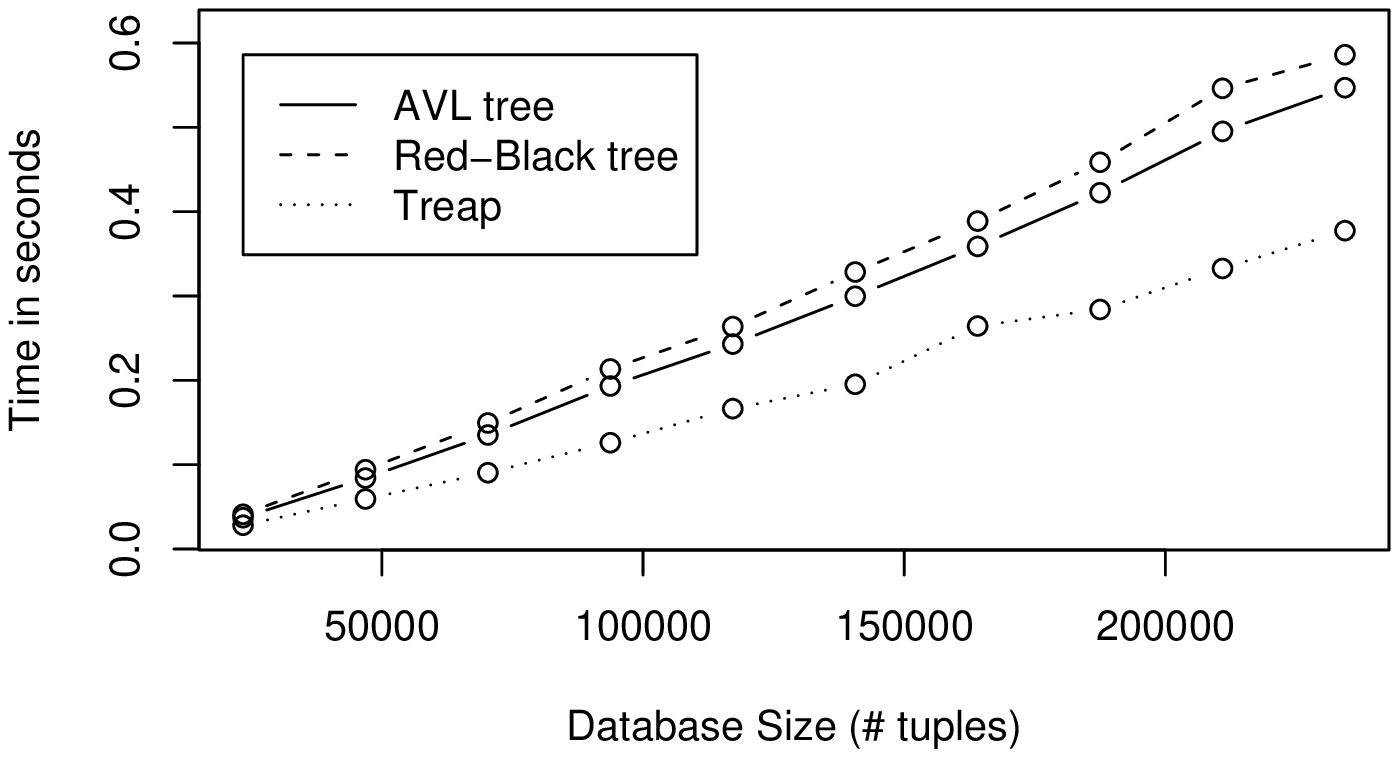}}
  \subfigure[RoE=1\%]{\includegraphics[width=\scale\columnwidth]{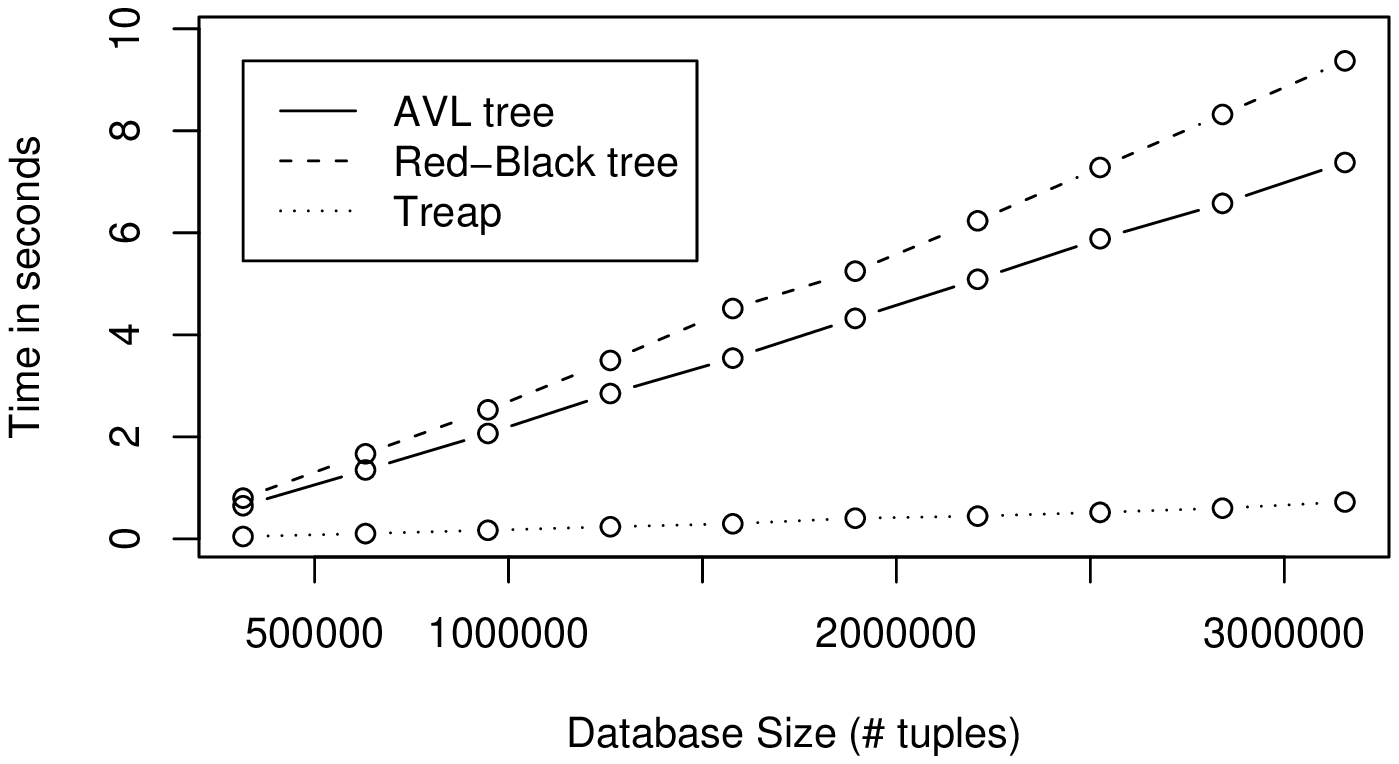}}
  \subfigure[RoE=100\%]{\includegraphics[width=\scale\columnwidth]{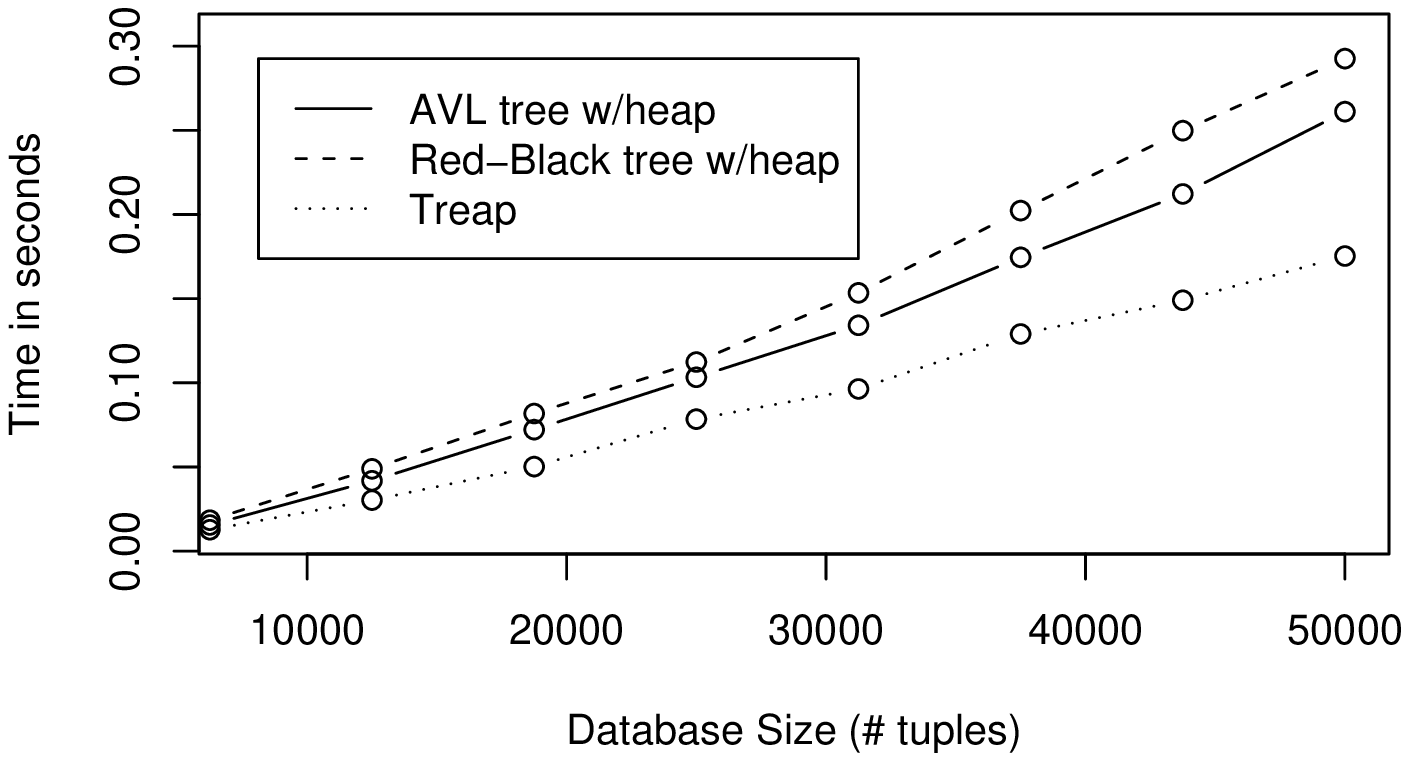}}
  \subfigure[RoE=5\%]{\includegraphics[width=\scale\columnwidth]{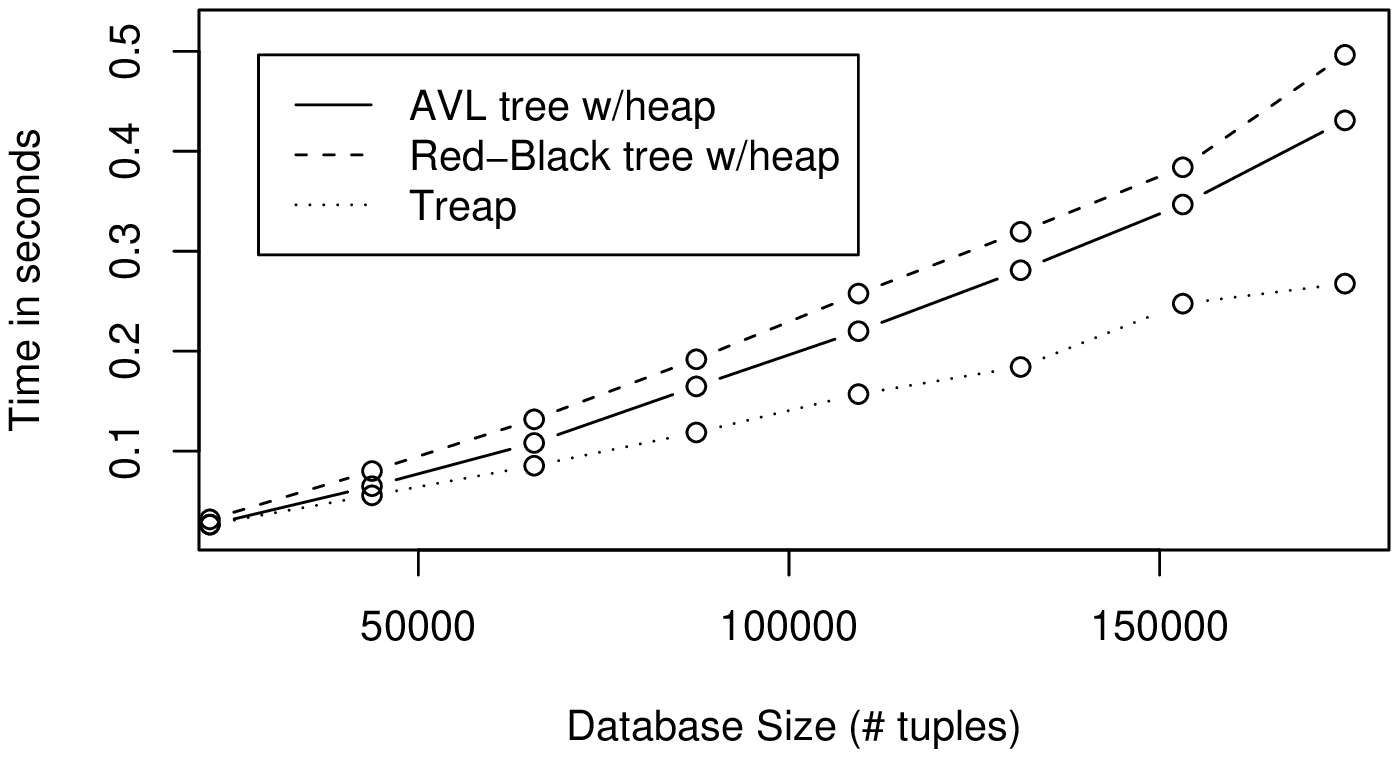}}
  \subfigure[RoE=1\%]{\includegraphics[width=\scale\columnwidth]{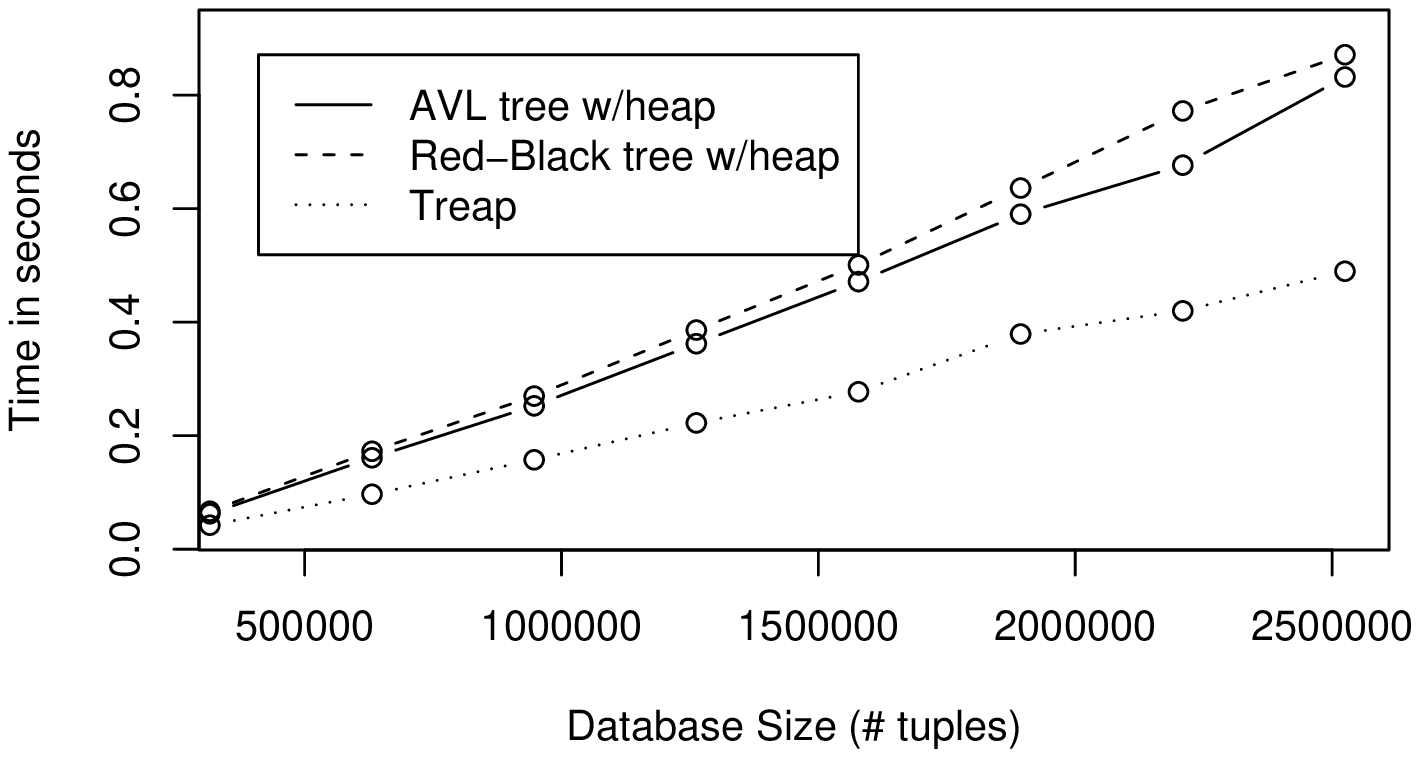}}
  \caption{Performance Comparison to AVL and Red-Black Trees with and without supporting heaps.}
  \label{fig:maintenance}
\end{figure*}

Figure~\ref{fig:maintenance} compares the cost of maintaining treaps
to the cost of maintaining the other well-known data structures which
were adapted to support expiration time.  The RoE varies from very
dynamic 100\% to much more static 1\%.
It turns out that treaps never perform significantly worse than the
other data structures but scale much better, both in terms of memory
requirements and processor time, for databases with relatively small
Rate of Expiration, which we consider a typical case.  In more detail,
Figures~\ref{fig:maintenance}(a)--~\ref{fig:maintenance}(c) illustrate
that treaps outperform AVL trees and Red-Black trees for RoEs of 5\%
and 1\%, whereas they incur only a small overhead for an RoE of
100\%.  Note that in this case expiration is done on AVL and Red-Black
using traversals at the end of each interval but no additional memory
is needed for supporting data structures; thus, expiration is not
eager.  However, eager expiration can be implemented with supporting
heaps as shown in
Figures~\ref{fig:maintenance}(d)--~\ref{fig:maintenance}(f). Note that
in these figures AVL and Red-Black trees are combined with heaps to
support eager expiration.  However, this incurs a memory overhead for
these data structures so that, given a fixed-size main memory, a treap
could index more than twice the data size than the competitors.
Nevertheless, treaps outperform the other data structures in all cases
although not as clearly as in the heap-less experiment.  

\subsubsection{Query Performance}
\label{sec:queryperformance}

This subsection considers the question what \emph{query} performance
(rather than the cost of maintenance) treaps feature in comparison to
AVL and Red-Black trees.  As mentioned earlier,
Figure~\ref{fig:queries} shows the performance for two important query
primitives used frequently in data management: traversals and lookup
queries.  It turns out that, as Figure~\ref{fig:lookups} shows, treaps
consistently outperform Red-Black trees and are \emph{en par} with AVL
trees with respect to point queries or lookups.  For small databases
AVL trees exhibit a slightly better performance where treaps are
slightly ahead for larger databases.  Similarly, for scanning the data
set in sort order, treaps perform slightly worse than both, Red-Black
and AVL trees, for small databases; for large databases, they are
again ahead of Red-Black trees as Figure~\ref{fig:traversal} shows.
However, the important point here is that the probabilistic
performance guarantees of treaps do not incur a significant (if at
all) penalty on query performance.

\subsection{Main Memory Performance}
\label{sec:caching}

We end the experimental study by covering some aspects of the
main-memory performance of persistent treaps.
We observe that the data structure does not exhibit locality, as do
B-trees~\cite{RM72} or T-trees~\cite{LC86}: updates, \ie~insertions,
deletions, and expiration, are very likely to affect a large portion
of the path from root to leaf level.  Additionally, to achieve
persistence, the node-copying method needs $O(\log n)$ memory per
update (but it also releases $O(\log n)$ memory as soon as all threads
pointing to the version terminate).  Thus, updates are expensive in
terms of cache utilisation: (1)~in-place pointer swizzling is
impossible because this would forestall persistence and thus
concurrency, and (2)~since each rotation requires a memory allocation,
we are very likely to get a cache miss (assuming that new memory is
rarely located inside any of the CPU caches).

Generic, cache-oblivious pagination for tree-shaped data
structures~\cite{BRJ02} are not applicable in our scenario since they
require the data set to be mostly static, whereas our setting is
highly dynamic.  However, there is one piece of good news. For
inserts, the nodes between the leaf level and the root that are
traversed when the heap property is being re-established are already
located in the CPU cache. As a result, no cache miss is likely to
occur during comparisons.

Treaps share an interesting performance feature with B-trees
with respect to bulk-loading.  Bulk-loading pre-sorted data,
\ie~data sorted according to the key attribute (not the expiration time),
turns out to be an order of magnitude cheaper in our experiments than
bulk-loading of unsorted data.  As with B-trees, this is due to caching
effects: when inserting a tuple from the root down to the leaf level,
the search path is most likely to be present in the level 1 and level 2
CPU caches.

To achieve predictable performance behaviour, we put great effort into
ensuring that the implementation does not allocate and release memory
more often than absolutely necessary.  In earlier implementations, the
creation of local objects in, for example, the insertion procedure
produced more garbage than necessary; this resulted in much larger
bandwidths than now displayed in Figures~\ref{fig:large_ana}
and~\ref{fig:small_ana}.  We found it to be of prime importance to
carefully analyse the memory allocation when implementing database
operators in garbage-collecting environments.  Still, we feel that the
advantages of using a garbage collector, such as easier memory
management, lower complexity of algorithms or data structures, and a
straight-forward implementation of persistence and concurrency,
outweigh the disadvantages, namely a diligent analysis of memory
allocation patterns.  Overall, a significant performance penalty can
be avoided by a diligent analysis of allocation patterns.
Thus, in our current implementation, the uncertainties caused by
\textit{amortised} costs and garbage collection add up but still allow
performance prediction with a degree of accuracy.  The overall
throughput of the data structure turns out to be fairly stable,
\ie~the overall execution times of, \eg~80,000 insertions do not
differ to a great deal.  Still, if we compare Figures~\ref{fig:large_ops}
and~\ref{fig:small_ops}, we see that the larger database, the harder
it becomes to predict performance.

\subsection{Further Issues}
\label{sec:extensions}

\begin{figure*}[!t]
  \newcommand{\scale}{0.49}
  \centering
  \subfigure[RoE=100\%]{\includegraphics[width=\scale\columnwidth]{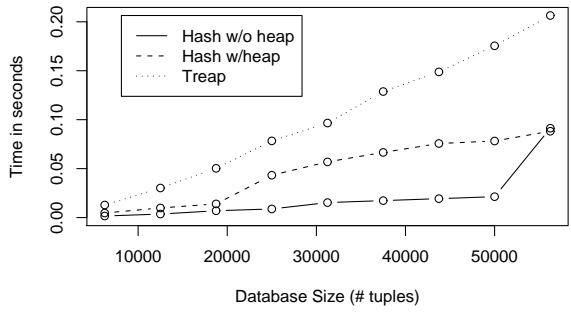}}
  \subfigure[RoE=5\%]{\includegraphics[width=\scale\columnwidth]{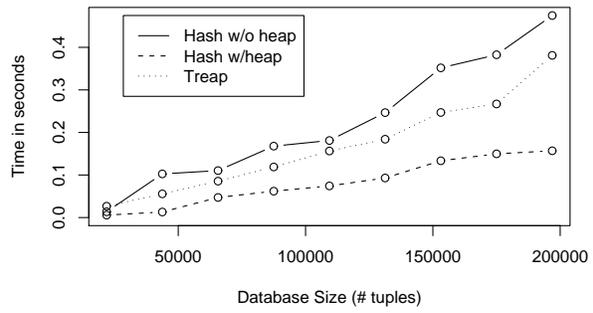}}
  \subfigure[RoE=1\%]{\includegraphics[width=\scale\columnwidth]{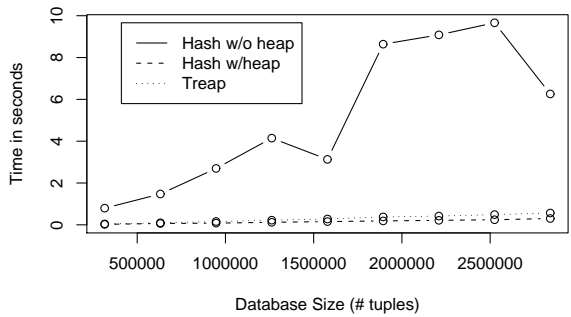}}
  \caption{Performance Comparison to hashtables with and without supporting heaps.}
  \label{fig:hashtreap}
\end{figure*}

For completeness, a few additional aspects are covered.

\subsubsection{Treaps and Non-Unique Indexes}

By altering the find and insert functions, it is possible to use the
data structure as a secondary index.  The find function now has to
return an iterator rather than a node.  It is necessary to check
whether the right and left children of a node $n$ have the same key as
$n$.  When we look for a key $k$, we have to look for its ancestor,
\ie~we effectively issue a range query.  The insert function must now
make sure that the tree does not lose balance when key duplicates are
inserted.  This is done by using a random-number generator and
executing random rotations within the set of duplicate keys.

\subsubsection{Non-Numerical Expiration Times}

Sometimes users are interested in tagging tuples with non-numerical
expiration times.  As mentioned in Section~\ref{sec:sumdis}, a natural
example would be infinity for an item which never expires, \ie~a tuple
carrying the usual SQL semantics which only ceases to be part of the
database if it as explicitly deleted.  To accommodate this feature it
seems easiest to use an extended comparison function $<^{exp}$ for the
expiration time in the following way ($T$ denotes the domain of
expiration times excluding $\infty$):
\[ x <^{exp} y=\left\{
    \begin{array}{ll}
      x < y                              & x,y\in T\\
      \textrm{true}                      & x\in T, y=\infty\\
      \textrm{false}                     & x=\infty, y\in T\\
      \textrm{true or false with } p=0.5 &\textrm{otherwise}\\
    \end{array} \right. \]

\noindent
The reason why this works is similar to the argument we presented for
secondary indexes, but this time the argument is applied to the
expiration time. To maintain the probabilistic balance of the tree, we
have to ensure that no correlation between the key and expiration
time exists.

\subsubsection{Robustness of Treaps}

The primary reason why treaps work well for data with expiration times
is that key values and expiration times are statistically independent.
Fill~\cite{Fil96} offers justifications for why this makes sense in
practise.  Seidel and Aragon~\cite{SA96} prove that the criterion of
statistical independence is sufficient, but also that it can be
relaxed significantly without adversely affecting performance. It
suffices if the key and the expiration time can be distinguished by
$\log n$ random bits, where $n$ is the number of tuples in the data
structure.

If, in an application, a correlation problem arises, we can use a hash
function $h(k)$ to ensure sufficient independence between keys and
expiration times.  This is akin to the idea presented by Seidel and
Aragon~\cite{SA96} of providing for sufficient statistical
independence between the key and priority (expiration time in our
case) by deriving the priority from the key with a hash function. This
way, the priority also becomes implicit in the key.  However, in our
setting priorities or expiration times are not just random values, but
carry actual meaning.  So applying this idea comes at the cost that
range queries on the key attributes can no longer be supported
efficiently.  However, a lookup query retains its original meaning
provided the lookup key $k$ is hashed to $h(k)$ before the lookup is
done.

Once we decide to use a hash function and, thus, deliberately do
without the opportunity to execute range queries efficiently or to
exploit sort order, hash tables become a competitor to treaps as
discussed and quantified in Figure~\ref{fig:hashtreap}.  The figures
show that a hash table with a supporting heap can outperform treaps at
the cost of increased memory since, again, two data structures have to
be maintained instead of only one.  On the negative side, the figures
also show that hash tables have much more trouble than treaps with
providing performance guarantees, as the erratic curve especially in
Figure~\ref{fig:hashtreap}(c) and the outlier in
Figure~\ref{fig:hashtreap}(a) show.

\subsubsection{Implementation Issues}

The balance of the treap as discussed above is critical not only in
terms of search performance, but also in terms of system stability. If
the treap becomes unbalanced, we are likely to produce a stack
overflow when we search the critical, unbalanced region of the tree.
This might be an error that is hard to detect, check for, and
prevent.

\subsection{Summary and Discussion}
\label{sec:sumdis}

In this section, we try to give a concise summary of the performance
evaluation of this paper.  (1)~Treaps as well as hash tables, AVL, and
Red-Black trees without supporting heaps have low memory
requirements in the sense that there is just one data structure to
contain all data.  For hash tables, AVL, and Red-Black trees this
comes at the cost that quality of service is an issue, \ie~the time
between logical expiration and physical removal can grow too large for
some applications; thus, these structures need a good and possibly
application-specific cleansing strategy.  Treaps do not have this
advantage and provide optimal quality of service.
(2)~The use of supporting heaps as priority queues for eager
expiration solves the quality of service issue but incurs a space
overhead needed to implement the additional structure.  It also makes
deletion \textit{very} expensive -- if not prohibitively expensive --
assuming that data should be deleted from both the index structure and
the priority, which in terms of memory requirements seems desirable in
a main-memory setting.
Allowing the heap to grow out of sync with the primary by removing
data only from the index might help but,
again, brings about a space penalty and the need for a heap cleansing
strategy.  In terms of implementation, it is noteworthy that
supporting heaps only contain \textit{finite} expiration times, if we
allow expiration times to be infinite as introduced in
Section~\ref{sec:extensions}.
(3)~Hash tables seem to have difficulty coping with the increased
dynamicity of the data which expiration incurs.  Adaptable and
application-specific growing and shrinking might help.  However, in
conjunction with supporting heaps they can outperform treaps by up to
one third at the cost of a space overhead of a factor of two.  Hash
tables are only a competitor if key/expiration correlation is an issue
and range queries are not required.

In summary, treaps are on par with competing data structures in most
cases and outperform them to various degrees in many interesting
cases, both in terms of maintenance cost and memory requirements.

\section{Related Work}
\label{sec:related}

At the level of query languages and data models, which is not the
focus of this paper, the concept of expiration time relates to the
concept of vacuuming~\cite{J95,SJM02}.  With vacuuming, it is possible
to specify rules that delete data: when the preconditions, \eg related
to time, in the head of a rule, are met, the data identified by the
body of the rule are logically deleted from the database.  Like
expiration time, vacuuming separates logical deletion from physical
deletion.  But whereas expiration times are explicitly associated with
the tuples in the database, vacuuming specifies which data to delete
in separate rules.  We believe that the techniques presented in this
paper may be relevant for the efficient implementation of time-based
vacuuming.  Stream databases~\cite{ABW03}, on the other hand,
allow users to specify query windows; in this sense, they take an
approach which is opposite to expiration times, which let the data
sources declare how long a tuple is to be considered current.
Some works that refer to the term ``expiration'' are slightly related
to expiration time and thus this paper's contribution.  Expiration has
been used in the context of view self-maintenance: Here the problem is
which data that can be removed (``expired'') without this affecting the
results of a predetermined set of queries (views)~\cite{GLY98,T03}.

The use of expiration time has been studied in the context of
disk-based indexing with the objective of supporting location-based
services~\cite{SJ02}. The idea is that locations reported by moving
objects that have not been updated explicitly for some time are
considered inaccurate and should thus be expired. The
R$^{\mathrm{EXP}}$-tree extends the R-tree to index the current and
anticipated future positions of two- and three-dimensional points,
where the points are given by linear functions of time and are
associated with expiration times.
We are not aware of any related
research on main memory based indexing that incorporates expiration
time.

Okasaki~\cite{Oka98} offers an excellent introduction to purely
functional, \ie~persistent, data structures.  Our primary data
structure, the (\textit{persistent}) \textit{treap}, is described and
analysed in substantial detail by Seidel and Aragon~\cite{SA96}; it
was first introduced by McCreight~\cite{McC85}.  Later, treaps were
primarily seen and interpreted as randomised search trees~\cite{AS89}.
Treaps have been used in a number of contexts, especially with random
values~\cite{MR98,Nil97}; however, we are not aware of any
time-related applications.  Heaps are a classical data structure in
computer science~\cite{Knu98}.

In this paper, we technically achieve concurrency on persistent treaps through
versioning~\cite{BHG87} by implementing the node-copying
method~\cite{DSD+89}.  In a database context, Lomet and
Salzberg~\cite{LS89,LS90} present versioning supporting variants of
B-trees and discuss related issues.  Finally, we remark that
distributed garbage collection also shares similarities with expiring
data in databases; especially eager collection is sensible when scarce
resources have to be freed up~\cite{PS95}.

\section{Conclusion and Future Work}
\label{sec:conclusion}

This paper argues that expiration time is an important concept for
data management in a variety of application areas, including heartbeat
patterns in mobile networks and short-lived data; it presents an
efficient main-memory data structure along with algorithms for data
with expiration time.  Through comprehensive and comparative
performance experiments, the paper demonstrates that the
implementation scales well beyond data volumes produced by current
mobile applications and thus is suited for advanced applications and
prototypes.

Data expiration is an important and natural concept in many volatile
application settings where traditional ACID semantics are not
appropriate.  Usually, devices such as mobile phones, PDA's, sensors,
and RFID tags are characterised by intermittent connectivity and often
do not need a full-blown transaction system for many tasks.  Thus,
data management applications can benefit from being expiration
time-enabled by experiencing a lower transaction workload and reduced
network traffic while at the same time being able to free memory
occupied by stale data immediately.  Additionally, it can also be
argued~\cite{SJ03} that expiration time is orthogonal to the notions
of transaction and valid time and can be supported in a temporal
database environment.  Expiration times also have the potential of
making for simpler application logic by removing the need for
``clean-up'' transactions.  The paper demonstrates through experiments
that a persistent treap, which is a binary tree with respect to a key
and a heap with respect to the expiration time, is an effective tool
for handling expiration times in main-memory settings.

Several interesting directions for \textit{future research} exist in
relation to the support for expiration time in data management.  When
data are not as short-lived as in our application, it might be
beneficial to develop strategies for extending standard
secondary-memory data structures, \eg~heap files, B-Trees, and Hash
files, with expiration time support.  We anticipate that expiration
for secondary-memory structures requires strategies different from
those presented in this paper.  Furthermore, to take full advantage of
DBMS technology, expiration times have to be sensibly integrated into
SQL's isolation levels and transaction system.

\section*{Acknowledgements}

The authors would like to thank Peter Schartner for helpful comments
and pointers as well as Laurynas Speicys, Simonas \v{S}altenis and
Kristian Torp for helpful discussions and their collaboration.

\bibliographystyle{plain}
\small{\textit{References containing URLs are valid as of 1 November 2004.}}

\end{document}